\documentclass[%
print,
superscriptaddress,
 amsmath,amssymb,
 aps,twocolumn
]{revtex4}

\usepackage{times}
\usepackage{graphicx}
\usepackage{dcolumn}
\usepackage{bm}
\usepackage{dsfont}
\usepackage{color}
\usepackage[landscape,papersize={297.1mm,210mm},left=1.9cm,right=1.6cm,top=2.7cm,bottom=2.8cm]{geometry}
\usepackage[                   
            pdfstartview=FitH,
            colorlinks, 
            pdfborder=001,   
           linkcolor=blue,
           anchorcolor=blue,
            citecolor=blue,
            urlcolor=blue
            ]{hyperref}
\usepackage{amsthm}
\newtheoremstyle{my}{}{}{\upshape}{\parindent}{\itshape}{.}{ }{}
\theoremstyle{my}
\newtheorem{theorem}{Theorem}
\newtheorem{lemma}{Lemma}
\newtheorem{proposition}{Proposition}
\newtheorem{definition}{Definition}

\newtheorem{Ob}{Observation}
\newtheorem*{Co}{Construction}

\usepackage{subfigure}
\usepackage{appendix}
\usepackage{tabularx}
\usepackage{array}
\usepackage{xcolor}
\usepackage{multirow,bigdelim}
\usepackage{blindtext}
\usepackage{multirow}
\usepackage{setspace}
\usepackage{etoolbox}
\let\bbordermatrix\bordermatrix
\patchcmd{\bbordermatrix}{8.75}{4.75}{}{}
\patchcmd{\bbordermatrix}{\left(}{\left[}{}{}
\patchcmd{\bbordermatrix}{\right)}{\right]}{}{}
\makeatletter

\newcommand{\Rmnum}[1]{\expandafter\@slowromancap\romannumeral #1@}

\makeatother
\begin{document}
\title{Strongest quantum nonlocality in $N$-partite systems}
\author{Mengying Hu}
\author{Ting Gao}
\email{gaoting@hebtu.edu.cn}
\affiliation{School of Mathematical Sciences, Hebei Normal University, Shijiazhuang 050024, China}
\affiliation{Hebei Mathematics Research Center,  Hebei Normal University, Shijiazhuang 050024, China}
\affiliation{Hebei International Joint Research Center for Mathematics and Interdisciplinary Science, Hebei Normal University, Shijiazhuang 050024, China}
\author{Fengli Yan}
\email{flyan@hebtu.edu.cn}
\affiliation{College of Physics, Hebei Key Laboratory of Photophysics Research and Application, Hebei Normal University, Shijiazhuang 050024, China}

\begin{abstract}
A set of orthogonal states possesses the strongest quantum nonlocality if only a trivial orthogonality-preserving positive operator-valued measure (POVM) can be performed for each bipartition of the subsystems. This concept originated from the strong quantum nonlocality  proposed by Halder $et~al.$ [\href{https://journals.aps.org/prl/abstract/10.1103/PhysRevLett.122.040403} {Phy. Rev. Lett. \textbf{122}, 040403 (2019)}], which is a stronger manifestation of nonlocality based on locally indistinguishability and finds more efficient applications in quantum information hiding. 
However, demonstrating the triviality of orthogonality-preserving local measurements (OPLMs) is not straightforward. In this paper, we present a sufficient and necessary condition for trivial OPLMs in $N$-partite systems under certain conditions. By using our proposed condition, we deduce the minimum size of set with the strongest nonlocality in system $(\mathbb{C}^{3})^{\otimes N}$, where the genuinely entangled sets constructed in Ref. [\href{https://journals.aps.org/pra/abstract/10.1103/PhysRevA.109.022220} {Phys. Rev. A \textbf{109}, 022220 (2024)}] achieve this value. As it is known that studying construction involving fewer states with strongest nonlocality contribute to reducing resource consumption in applications. Furthermore, we construct strongest nonlocal genuinely entangled sets in system $(\mathbb{C}^{d})^{\otimes N}~(d\geq4)$, which have a smaller size than the existing strongest nonlocal genuinely entangled sets as $N$ increases. Consequently, our results contribute to a better understanding of strongest nonlocality.
\end{abstract}

\pacs{ 03.67.Mn, 03.65.Ud, 03.67.-a}

\maketitle

\section{Introduction}
Locally indistinguishable states exhibit quantum nonlocality, which means that a set of orthogonal quantum states cannot be discriminated by local operations and classical communication (LOCC). This seminal concept, introduced by Bennett $et~al$. \cite{PRA.59.1070}, has been studied extensively \cite{PRL.85.4972,PRL.89.147901,PRA.75.014305,PRA.75. 052313,PRA.74.052103,ieee.4957660,PRA.84.012304,PRA.90.022313,PRA.92.032313,QIP.10.1007,PRA.98.022303,
PRL.87.277902,PRA.70.022304,PRL.92.177905,NJP.13.123013,PRL.109.020506,PRA.99.022307,PRL.126.210505}. This kind of state plays a crucial role in various quantum information processing tasks, such as accessing, transmitting, and storing quantum information.
When information is encoded in locally indistinguishable states, it cannot be locally retrieved by spatially separated users who are restricted to using only LOCC, unless some (all) are together. As a consequence, these states have found practical applications in data hiding \cite{PRL.86.5807,ieee.48.580} and quantum secret sharing \cite{PRA.91.022330,PRA.95.022320}.

Local irreducibility, a stronger concept than locally indistinguishability, refers to a set of orthogonal states that cannot be eliminated one or more states by orthogonality-preserving local measurements (OPLMs), and was proposed by Halder $et~al$.\;\cite{PRL.122.040403}. It induces strong quantum nonlocality. A set of orthogonal states is strongly nonlocal if it is locally irreducible in every bipartition of the subsystems \cite{PRL.122.040403}. Naturally, information hidden in strongly nonlocal states enhances security since no information can be obtained even if $N-1$ users are collusive  \cite{PRA.105.022209}, indicating the higher practical value of such states. By definition, proving strong nonlocality bases on demonstrating local irreducibility. However, irreducibility is not easy to determine. At present, lots of efforts have been devoted to the construction of different types of strongly nonlocal sets \cite{
PRA.99.062108,PRA.102.042228,Quantum.6.619,PRA.106.052209,PRA.107.042214,
Quantum.8.1649,PRA.102.042202,PRA.104.012424,PRA.105.022209,Quantum.7.1101,
PRA.108.062407,PRA.109.022220, PRA.109.052422,PRA.109.052211,arXiv: 2403.10969}. The predominant approach employed by these papers \cite{PRA.99.062108,PRA.102.042228,Quantum.6.619,PRA.106.052209,PRA.107.042214,
Quantum.8.1649,PRA.102.042202,PRA.104.012424,PRA.105.022209,Quantum.7.1101,
PRA.108.062407,PRA.109.022220,PRA.109.052422,PRA.109.052211} is to prove mathematically that the OPLMs can only be trivial, which is also a sufficient condition for local irreducibility. This is due to that a trivial measurement (a measurement $\{E_{x}\}_{x}$ is called trivial if all the positive semidefinite operators are proportional to the identity operator) is orthogonality-preserving and gives us no information.

Wang $et~al$. \cite{PRA.104.012424} defined that a set possesses the strongest nonlocality if only a trivial orthogonality-preserving positive operator-valued measure (POVM) can be performed for each bipartition of the subsystems. Therefore, the strongly nonlocal sets \cite{
PRA.99.062108,PRA.102.042228,Quantum.6.619,PRA.106.052209,PRA.107.042214,
Quantum.8.1649,PRA.102.042202,PRA.104.012424,PRA.105.022209,Quantum.7.1101,
PRA.108.062407,PRA.109.022220,PRA.109.052422} constructed using the ``trivial OPLMs'' technique have the property of  the strongest nonlocality. However, as the number and dimension of subsystems increase in multipartite systems, proving the triviality of OPLM becomes more intricate. Shi $et~al$.\;\cite{Quantum.6.619} developed two lemmas, Block Zeros Lemma and Block Trivial Lemma, which contribute to simplifying the proof process. To further alleviate this difficulty, we firstly present several sufficient conditions for Block Trivial Lemma. Subsequently, under certain conditions, we present a sufficient and necessary condition for the triviality of OPLM in every bipartition in $N$-partite systems.  By utilizing this result, we prove the minimum size of strongest nonlocal set in system $(\mathbb{C}^{3})^{\otimes N}$ under certain conditions, which is reached by the genuinely entangled sets constructed in Ref. \cite{PRA.109.022220}.

Seeking fewer states that exhibit the strongest nonlocality contributes to resource conservation in quantum information processing tasks.  Much efforts have been made on constructing smaller set for both orthogonal product states \cite {PRA.99.062108,PRA.102.042228,Quantum.6.619,PRA.106.052209,PRA.107.042214,Quantum.8.1649} and entangled states \cite{PRA.102.042202,PRA.104.012424,PRA.105.022209,Quantum.7.1101,PRA.108.062407,PRA.109.022220}.
Intuitively, entanglement gives rise to nonlocality as entangled states \cite{RMP.81.865,Physics.1.195,PRL.23.880,PRA.83.022319,RMP.86.839,PRL.112.180501,CHB.24.070301,IJTP.55.4231,PRA.98.062103} exhibit Bell nonlocality \cite{RMP.86.839} by violating Bell inequality.  Bennett $et~al$. \cite{PRA.59.1070} presented an indistinguishable orthogonal product basis (OPB) in Hilbert space $\mathbb{C}^{3}\otimes \mathbb{C}^{3}$ demonstrating that local indistinguishability is not directly related to entanglement. Subsequently, Walgate $et~al$.\;\cite{PRL.85.4972} showed any pair of orthogonal states can always be perfectly distinguished locally, whether entangled or not. This indicates that entanglement is not a necessary condition for nonlocality.

Recently, a lower bound for the strongest nonlocal set on multipartite systems was proved by Li $et~al$.\;\cite{Quantum.7.1101}.
In system $\otimes^{N}_{i=1}\mathbb{C}^{d_{i}}$ $(d_{i}$ is the dimension of the $i$th subsystem), it has been proven that the size of the strongest nonlocal sets can not be smaller than $\mathop{\mathrm{max}}\limits_{i}\{\hat{d}_{i}+1\}$, where $\hat{d}_{i}=(\prod\limits_{j=1}^{N}d_{j})/d_{i}$. In tripartite systems, Zhen $et~al$.\;\cite {PRA.109.052422} constructed the sets that reach the lower bound, one of which is product state and the rest is entangled state. So natural questions arise:
What role do the entangled states play in the manifestation of the strongest nonlocality?
How smaller can a strongest nonlocal entangled set could be in some given system?
In the existing research on entangled states \cite{PRA.102.042202,PRA.104.012424,PRA.105.022209,Quantum.7.1101,PRA.108.062407,PRA.109.022220}, Li $et~al$. \cite{Quantum.7.1101} constructed the strongest nonlocal orthogonal genuinely entangled sets (OGESs) with size $\prod\limits_{i=1}^{N}d_{i}-\prod\limits_{i=1}^{N}(d_{i}-1)+1$ in general multipartite systems $\otimes^{N}_{i=1}\mathbb{C}^{d_{i}}$. Then, in system $(\mathbb{C}^{3})^{\otimes N}$, Hu $et~al.$ \cite{PRA.109.022220} presented a more less set of OGESs possessing the strongest nonlocality. Do the smaller set of OGESs exist in high-dimensional multipartite systems?  In this paper, we generalize the set in Ref.\;\cite{PRA.109.022220} to system $(\mathbb{C}^{d})^{\otimes N}$ and demonstrate that the size in our construction is much smaller than that of Ref.\;\cite{Quantum.7.1101} as $N$ increases while keeping $d$ fixed. It also provides an answer to
the question given in Ref.\;\cite{PRA.105.022209}, ``How do we construct a strongly
nonlocal orthogonal genuinely entangled set in $(\mathbb{C}^{d})^{\otimes N}$
for any $d\geq2$ and $N\geq5$?"

The paper is organized as follows. In Sec. \ref{Q1}, some necessary notations and definitions are introduced. In Sec. \ref{Q2}, we present a sufficient and necessary condition for the triviality of local orthogonality-preserving POVM under bipartition in $N$-partite systems. In this condition, we discuss the minimum size of set in system $(\mathbb{C}^{3})^{\otimes N}$. In Sec. \ref{Q3}, we construct strongest nonlocal OGESs in system $(\mathbb{C}^{d})^{\otimes N}~(d\geq4,N\geq3)$. Finally, we draw a conclusion in Sec. \ref{Q4}.

\section{Preliminaries}\label{Q1}
We first introduce some definitions and lemmas needed in this paper. A POVM is a set of semidefinite operators $\{E_{m}=M^{\dagger}_{m}M_{m}\}$ such that $\sum_{m}E_{m}=\mathbb{I}$, where $\mathbb{I}$ is identity operation. Throughout this paper, we consider only pure state and POVM measurements, and do not normalize the states for simplicity.

Assume that $E=(a_{i,j})_{i,j\in\mathbb{Z}_{d}}$ is the matrix representation of the operator $E=M^{\dag}M$ under the bases $\mathcal{B}:=\{|0\rangle,|1\rangle,\cdots,|d-1\rangle\}$. Given two nonempty subsets $\mathcal{S}$ and $\mathcal{T}$ of $\mathcal{B}$, where $s=|\mathcal{S}|$, $t=|\mathcal{T}|$ are the cardinality of $\mathcal{S}$ and $\mathcal{T}$, respectively.
$_{\mathcal{S}}E_{\mathcal{T}}:=\sum\limits_{|i\rangle\in \mathcal{S}}\sum\limits_{|j\rangle\in\mathcal{T}}a_{i,j}|i\rangle\langle j|$ is a sub-matrix of $E$ with row coordinates $\mathcal{S}$ and column coordinates $\mathcal{T}$, specifically, $_{\mathcal{S}}E_{\mathcal{S}}$ is denoted by $E_{\mathcal{S}}$. Ref. \cite{Quantum.6.619} gives two lemmas.
\begin{lemma}[Block~Zeros~Lemma]\label{le-zeros}
Suppose that $\{|\psi_{i}\rangle\}^{s-1}_{i=0}$ and $\{|\phi_{j}\rangle\}^{t-1}_{j=0}$ are two orthogonal sets spanned by $\mathcal{S}$ and $\mathcal{T}$ respectively, if $\langle\psi_{i}|E|\phi_{j}\rangle=0$ for any $i\in\mathbb{Z}_{s}$, $j\in\mathbb{Z}_{t}$, then $_{\mathcal{S}}E_{\mathcal{T}}=\mathbf{0}$ and $_{\mathcal{T}}E_{\mathcal{S}}=\mathbf{0}$.
\end{lemma}
\begin{lemma}[Block~Trivial~Lemma]\label{le-trivial}
Let $\{|\psi_{j}\rangle\}^{s-1}_{j=0}$ be an orthogonal set spanned by $\mathcal{S}$, assume that $\langle\psi_{i}|E|\psi_{j}\rangle=0$ for any $i\neq j\in \mathbb{Z}_{s}$. If there exists a state $|u_{0}\rangle$ in $\mathcal{S}$ such that $_{\{|u_{0}\rangle\}}E_{S\setminus\{|u_{0}\rangle\}}=0$ and $\langle u_{0}|\psi_{j}\rangle\neq0$ for any $j\in \mathbb{Z}_{s}$, then $E_{\mathcal{S}}\propto \mathbb{I}_{\mathcal{S}}$.
\end{lemma}
\begin{lemma}(Shi $et~al$.\;\cite{Quantum.6.619})\label{le-est}
Let $\{|\psi_{j}\rangle\}$ be a set of orthogonal states in multipartite system $\otimes^{N}_{i=1}\mathbb{C}^{d_{i}}$. For each $i=1,2,\cdots,N$, define $\bar{A}_{i}=\{A_{1}A_{2}\cdots A_{N}\}\backslash\{A_{i}\}$ is the joint party of all but the $i$th party. If any OPLM on $\bar{A}_{i}$ is trivial, then the set $\{|\psi_{j}\rangle\}$ possesses the strongest nonlocality.
\end{lemma}
Consider an $N$-partite quantum system $\mathcal{H}=\mathcal{H}_{A_{1}}\otimes \mathcal{H}_{A_{2}}\otimes\cdots\otimes \mathcal{H}_{A_{N}}$ with dimensional $d_{k}$ for the $k$th $(1\leq k\leq N)$ subsystem.
The computational basis of the whole quantum system
is $\mathcal{B}=\{|i_{1}\rangle|i_{2}\rangle\cdots |i_{N}\rangle\mid i_{k}\in\mathbb{Z}_{d_{k}}\}$.  Let
\begin{equation}\label{eq:B}
\begin{aligned}
\mathcal{B}_{r}:= \{|r_{1}^{t}\rangle|r_{2}^{t}\rangle\cdots|r_{N}^{t}\rangle|r_{k}^{t}\in\mathbb{Z}_{d_{k}}\}_{t\in\mathbb{Z}_{|\mathcal{B}_{r}|}},~~r\in \mathcal{Q}
\end{aligned}
\end{equation}
are the disjoint subsets of $\mathcal{B}$, where $\mathcal{Q}=\{1,2,\cdots,q\}$.
We define the set of orthogonal states spanned by $\mathcal{B}_{r}$ as
\begin{equation}\label{eq:ss}
\begin{aligned}
\mathcal{S}_{r}:=&\big\{
|\Psi_{r,c}\rangle:=\sum\limits_{t\in\mathbb{Z}_{|\mathcal{B}_{r}|}}\omega^{c \cdot t}_{|\mathcal{B}_{r}|}|r_{1}^{t}\rangle|r_{2}^{t}\rangle\cdots|r_{N}^{t}\rangle\big\},
\end{aligned}
\end{equation}
where $\omega_{|\mathcal{B}_{r}|}=\mathrm{e}^{\frac{2\pi\sqrt{-1}}{{|\mathcal{B}_{r}|}}}$, $c\in\mathbb{Z}_{|\mathcal{B}_{r}|}$.

The set of $N$-tuples that corresponding to the basis vectors in $\mathcal{B}_{r}$ denotes as
\begin{equation}\label{eq:GN1}
\begin{aligned}
\mathcal{G}^{N}_{r}=\{(r_{1}^{t},r_{2}^{t},\cdots,r_{N}^{t})\}_{t\in\mathbb{Z}_{|\mathcal{B}_{r}|}}.
\end{aligned}
\end{equation}
We divide $\mathbb{Z}_{|\mathcal{B}_{r}|}$ into $m^{r}~(0< m^{r}\leq|\mathcal{B}_{r}|)$ disjoint subsets $R_{\tau}$ ($\tau\in \mathbb{Z}_{m^{r}}$), i.e., $\bigcup_{\tau\in \mathbb{Z}_{m^{r}}}R_{\tau}=\mathbb{Z}_{|\mathcal{B}_{r}|}$, for any $t\neq t^{'}\in R_{\tau}$, $r_{1}^{t}=r_{1}^{t^{'}}$. Let $r_{1}^{R_{\tau}}=r_{1}^{t}~(t\in R_{\tau})$, the set of $N$-tuples in $\mathcal{G}^{N}_{r}$ whose first component is same as $r_{1}^{R_{\tau}}$  is represented as
\begin{equation}
\begin{aligned}
\{r_{1}^{R_{\tau}}\}\times\{(r_{2}^{t},\cdots, r_{N}^{t})\}_{t\in R_{\tau}}=\{r_{1}^{R_{\tau}}\}\times \mathcal{G}^{N-1}_{[r,R_{\tau}]},
\end{aligned}
\end{equation}
where $\mathcal{G}^{N-1}_{[r,R_{\tau}]}=\bigcup_{t\in R_{\tau}}\{(r_{2}^{t},\cdots, r_{N}^{t})\}$. Therefore
\begin{equation}\label{eq:GNr}
\begin{aligned}
\mathcal{G}^{N}_{r}=\bigcup_{\tau\in \mathbb{Z}_{m^{r}}}\big(\{r_{1}^{R_{\tau}}\}\times\mathcal{G}^{N-1}_{[r,R_{\tau}]}\big).
\end{aligned}
\end{equation}

\begin{definition} $\mathcal{I}_{[r,R_{\tau}]}^{N-1}=\bigcup_{v\in\mathcal{O}}\mathcal{G}^{N-1}_{[v,V_{\gamma}]}~(r\notin\mathcal{O}\subset\mathcal{Q},~\gamma\in \mathbb{Z}_{m^{v}})$ is called the block inclusion (BI) set of $\mathcal{G}^{N-1}_{[r,R_{\tau}]}$, if for any $u\neq w\in \mathcal{O}$ there exist $\alpha\in \mathbb{Z}_{m^{u}}$, $\beta\in \mathbb{Z}_{m^{w}}$ such that $u_{1}^{U_{\alpha}}=w_{1}^{W_{\beta}}\neq r_{1}^{R_{\tau}}$  and  $\mathcal{G}^{N-1}_{[r,R_{\tau}]}\subset \mathcal{I}_{[r,R_{\tau}]}^{N-1}$. Specifically, $\mathcal{I}_{[r,R_{\tau}]}^{N-1}$ is called a tight block inclusion (TBI) set if there exists a subset $\mathcal{G}^{N-1}_{[x,X_{\lambda}]}\subset \mathcal{I}_{[r,R_{\tau}]}^{N-1}~(x\in \mathcal{O},~ \lambda\in \mathbb{Z}_{m^{x}})$  such that $|\mathcal{G}^{N-1}_{[r,R_{\tau}]}\bigcap\mathcal{G}^{N-1}_{[x,X_{\lambda}]}|=1$, and $\mathcal{G}^{N-1}_{[x,X_{\lambda}]}$ is called a tight sub-block.
\end{definition}
\begin{definition}
A family of sets $\{\bigcup_{\tau}\mathcal{G}^{N-1}_{[r,R_{\tau}]}\}_{r\in\mathcal{Q}}$ is connected if it cannot be divided into two groups of sets $\{\bigcup_{\gamma}\mathcal{G}^{N-1}_{[v,V_{\gamma}]}\}_{v\in\mathcal{O}}~(\mathcal{O}\subsetneqq \mathcal{Q})$ and $\{\bigcup_{\lambda}\mathcal{G}^{N-1}_{[x,X_{\lambda}]}\}_{x\in \mathcal{Q}\backslash\mathcal{O}}$ such that
\begin{equation}
\begin{aligned}
\Bigg(\bigcup_{v\in\mathcal{O}}\bigcup_{\gamma\in \mathbb{Z}_{m^{v}}}\mathcal{G}^{N-1}_{[v,V_{\gamma}]}\Bigg)\bigcap\Bigg(\bigcup_{x\in\mathcal{Q}\backslash\mathcal{O}}\bigcup_{\lambda\in \mathbb{Z}_{m^{x}}}\mathcal{G}^{N-1}_{[x,X_{\lambda}]}\Bigg)=\emptyset.
\end{aligned}
\end{equation}
\end{definition}
\section{THE SUFFICIENT AND NECESSARY CONDITION FOR THE TRIVIALITY
OF ORTHOGONALITY-PRESERVING LOCAL MEASUREMENTS AND THE SMALLEST SIZE OF STATES UNDER THIS CONSTRAINTS}\label{Q2}
According to Lemma~\ref{le-est}, to prove the strongest nonlocality, we need to show that the OPLM on $\bar{A}_{k}=\{{A}_{k+1}\cdots {A}_{N}\cdots {A}_{k-1}\}$ can only be trivial. In this section, we first give a sufficient condition for Lemma ~\ref{le-trivial}, then we present a sufficient and necessary condition for the triviality of OPLM.
Under this condition, we discuss the smallest size of set with strongest nonlocality in $(\mathbb{C}^{3})^{\otimes N}~(N\geq3)$.

Let $\{\Pi\}$ be the orthogonality-preserving POVM on subsystem $\bar{A}_{k}$, and $E=\mathbb{I}\otimes\Pi$.
For the states $\{\mathcal{S}_{r}\}$ in Eq. (\ref{eq:ss}), we have
\begin{equation}\label{eq:rs}
\langle\Psi_{r,c}|\mathbb{I}\otimes\Pi|\Psi_{r,c'}\rangle=0,~~~\langle\Psi_{r,c}|\mathbb{I}\otimes\Pi|\Psi_{v,c''}\rangle=0
\end{equation}
where $c\neq c'\in\mathbb{Z}_{|\mathcal{B}_{r}|}$, $c''\in\mathbb{Z}_{|\mathcal{B}_{v}|}$ $(r\neq v\in \mathcal{Q})$.
\begin{theorem}\label{th-1}
Consider the set $\mathcal{B}_{r}$ (\ref{eq:B}), if one of the following conditions is true for the set given by Eq. (\ref{eq:GNr}), Block Trivial Lemma (Lemma \ref{le-trivial}) was satisfied, i.e. $E_{\mathcal{B}_{r}}\propto \mathbb{I}_{\mathcal{B}_{r}}$.

(i) There exists $\tau_{0}\in \mathbb{Z}_{m^{r}}$ such that $|\mathcal{G}^{N-1}_{[r,R_{\tau_{0}}]}|=1$.

(ii) There is a set $\mathcal{G}^{N-1}_{[r,R_{\tau_{1}}]}~(\tau_{1}\in \mathbb{Z}_{m^{r}})$ has a TBI set.

(iii) There is a set $\mathcal{G}^{N-1}_{[r,R_{\tau_{2}}]}~(\tau_{2}\in \mathbb{Z}_{m^{r}})$ which has a BI set $\mathcal{I}_{[r,R_{\tau_{2}}]}^{N-1}= \bigcup_{v\in\mathcal{O}}\mathcal{G}^{N-1}_{[v,V_{\gamma}]}~(r\notin\mathcal{O}\subset\mathcal{Q})$, for each $v\in\mathcal{O}$, $\mathcal{G}^{N}_{v}$ satisfies condition (i) or (ii).
\end{theorem}
$\textit{Proof.}$
Condition (i) :
Consider the $N$-tuple $(r_{k}^{R_{\tau_{0}}},r^{t}_{k+1},\cdots,r^{t}_{N},\cdots,r^{t}_{k-1})$ in set $\{r_{k}^{R_{\tau_{0}}}\}\times\mathcal{G}^{N-1}_{[r,R_{\tau_{0}}]}$, its corresponding basis vector $|\textbf{\emph{r}}\rangle=|r_{k}^{R_{\tau_{0}}},r^{t}_{k+1},\cdots,r^{t}_{N},\cdots,r^{t}_{k-1}\rangle$ belongs to $\mathcal{B}_{r}$, for any state $|\textbf{\emph{r}}'\rangle=|r^{t'}_{k},r^{t'}_{k+1},\cdots,r^{t'}_{N},\cdots,r^{t'}_{k-1}\rangle\in
\mathcal{B}_{r}\backslash\{|\textbf{\emph{r}}\rangle\}$, there is $t'\notin R_{\tau_{0}}$ and $r^{t'}_{k}\neq r_{k}^{R_{\tau_{0}}}$, thus
\begin{equation}
\begin{aligned}
&\langle \textbf{\emph{r}}|E|\textbf{\emph{r}}'\rangle
=\langle \textbf{\emph{r}}|\mathbb{I}\otimes\Pi|\textbf{\emph{r}}'\rangle
=0,
\end{aligned}
\end{equation}
we get $_{\{|\textbf{\emph{r}}\rangle\}}E_{\mathcal{B}_{r}\backslash\{|\textbf{\emph{r}}\rangle\}}=0$, since $\langle \textbf{\emph{r}}|\Psi_{r,s}\rangle\neq0$, Lemma \ref{le-trivial} is satisfied.

Condition (ii): For the TBI set $\mathcal{I}_{[r,R_{\tau_{1}}]}^{N-1}=\bigcup_{v\in\mathcal{O}}\mathcal{G}^{N-1}_{[v,V_{\gamma}]}~(r\notin\mathcal{O}\subset\mathcal{Q},~\gamma\in \mathbb{Z}_{m^{v}})$, assume that $\mathcal{G}^{N-1}_{[x,X_{\lambda}]}\subset\mathcal{I}_{[r,R_{\tau_{1}}]}^{N-1}$ is a tight sub-block of set $\mathcal{G}^{N-1}_{[r,R_{\tau_{1}}]}$ and $\mathcal{G}^{N-1}_{[r,R_{\tau_{1}}]}\bigcap\mathcal{G}^{N-1}_{[x,X_{\lambda}]}=(x_{k+1},\cdots,x_{N},\cdots,x_{k-1})$, we can get $\textbf{\emph{r}}=(r^{R_{\tau_{1}}}_{k},x_{k+1},\cdots,x_{N},\cdots,x_{k-1})\in\{r^{R_{\tau_{1}}}_{k}\}\times\mathcal{G}^{N-1}_{[r,R_{\tau_{1}}]}$. For any state $|\textbf{\emph{r}}'\rangle=|r^{t'}_{k},r^{t'}_{k+1},\cdots,r^{t'}_{N},\cdots,r^{t'}_{k-1}\rangle\in\mathcal{B}_{r}\backslash\{|\textbf{\emph{r}}\rangle\}$,
if ${t'}\notin R_{\tau_{1}}$, i.e., $r^{t'}_{k}\neq r^{R_{\tau_{1}}}_{k}$, we have $\langle \textbf{\emph{r}}|E|\textbf{\emph{r}}'\rangle=\langle \textbf{\emph{r}}|\mathbb{I}\otimes\Pi|\textbf{\emph{r}}'\rangle=0$.
If ${t'}\in R_{\tau_{1}}$, we obtain that $(r^{t'}_{k+1},\cdots,r^{t'}_{N},\cdots,r^{t'}_{k-1})\in\mathcal{G}^{N-1}_{[r,R_{\tau_{1}}]}$, since $\mathcal{G}^{N-1}_{[x,X_{\lambda}]}$ is a tight sub-block,  there must have $u\neq x \in \mathcal{O}$ such that $(r^{t'}_{k+1},\cdots,r^{t'}_{N},\cdots,r^{t'}_{k-1})\in\mathcal{G}^{N-1}_{[u,U_{\alpha}]}\subset I^{N-1}_{[r,R_{\tau_{1}}]}$. Because  $|\textbf{\emph{x}}\rangle=|x^{X_{\lambda}}_{k},x_{k+1},\cdots,x_{N},\cdots,x_{k-1}\rangle\in \mathcal{B}_{x}$,  $|\textbf{\emph{u}}\rangle=|u^{U_{\alpha}}_{k},r^{t'}_{k+1},\cdots,r^{t'}_{N},\cdots,r^{t'}_{k-1}\rangle\in \mathcal{B}_{u}$ and $x^{X_{\lambda}}_{k}=u^{U_{\alpha}}_{k}$, it follows from Eq. (\ref{eq:rs}) that
\begin{equation*}
\begin{aligned}
\langle &\textbf{\emph{x}}|E|\textbf{\emph{u}}\rangle
\\
=&\langle x_{k+1},\cdots,x_{N},\cdots,x_{k-1}|\Pi|r^{t'}_{k+1},\cdots,r^{t'}_{N},\cdots,r^{t'}_{k-1}\rangle\\
=&0.
\end{aligned}
\end{equation*}
Since $r^{t'}_{k}= r^{R_{\tau_{1}}}_{k}$, we have
\begin{equation*}
\begin{aligned}
&\langle\textbf{\emph{r}}|E|\textbf{\emph{r}}'\rangle\\
=&\langle x_{k+1},\cdots,x_{N},\cdots,x_{k-1}|\Pi|r^{t'}_{k+1},\cdots,r^{t'}_{N},\cdots,r^{t'}_{k-1}\rangle\\
=&0,
\end{aligned}
\end{equation*}
meanwhile, $\langle \textbf{\emph{r}}|\Psi_{r,s}\rangle\neq0$, Lemma \ref{le-trivial} is satisfied.

Condition (iii):
For arbitrary two states $|\textbf{\emph{r}}\rangle=|r_{k}^{R_{\tau_{2}}},r^{t}_{k+1},\cdots,r^{t}_{N},\cdots,r^{t}_{k-1}\rangle\in\mathcal{B}_{r}$ and  $|\textbf{\emph{r}}'\rangle=|r^{t'}_{k},r^{t'}_{k+1},\cdots,r^{t'}_{N},\cdots,r^{t'}_{k-1}\rangle\in\mathcal{B}_{r}\backslash\{|\textbf{\emph{r}}\rangle\}$,
if $t'\notin R_{\tau_{2}}$, apparently, $\langle \textbf{\emph{r}}|E|\textbf{\emph{r}}'\rangle=0$. If $t'\in R_{\tau_{2}}$, we get $(r^{t}_{k+1},\cdots,r^{t}_{N},\cdots,r^{t}_{k-1})\neq(r^{t'}_{k+1},\cdots,r^{t'}_{N},\cdots,r^{t'}_{k-1})\in\mathcal{G}^{N-1}_{[r,R_{\tau_{2}}]}$. Since $\mathcal{G}^{N-1}_{[r,R_{\tau_{2}}]}$ has a BI set $\bigcup_{v\in\mathcal{O}}\mathcal{G}^{N-1}_{[v,V_{\gamma}]}$, suppose $(r^{t}_{k+1},\cdots,r^{t}_{N},\cdots,r^{t}_{k-1}),~(r^{t'}_{k+1},\cdots,r^{t'}_{N},\cdots,r^{t'}_{k-1})\in\mathcal{G}^{N-1}_{[v,V_{\gamma}]}~(v\in \mathcal{O})$, as $\mathcal{G}^{N}_{v}$ satisfies condition (i) or (ii), we have $E_{\mathcal{B}_{v}}\propto \mathbb{I}_{\mathcal{B}_{v}}$, i.e., for $|\textbf{\emph{v}}\rangle=|v^{V_{\gamma}}_{k},r^{t}_{k+1},\cdots,r^{t}_{N},\cdots,r^{t}_{k-1}\rangle$ and $|\textbf{\emph{v}}'\rangle=|v^{V_{\gamma}}_{k},r^{t'}_{k+1},\cdots,r^{t'}_{N},\cdots,r^{t'}_{k-1}\rangle$, there is
\begin{equation}
\begin{aligned}
&\langle \textbf{\emph{v}}|E|\textbf{\emph{v}}'\rangle\\
=&\langle r^{t}_{k+1},\cdots,r^{t}_{N},\cdots,r^{t}_{k-1}|\Pi|r^{t'}_{k+1},\cdots,r^{t'}_{N},\cdots,r^{t'}_{k-1}\rangle\\
=&0.
\end{aligned}
\end{equation}
Thus
\begin{equation}\label{}
\begin{aligned}
&\langle\textbf{\emph{r}}|E|\textbf{\emph{r}}'\rangle\\
=&\langle r^{t}_{k+1},\cdots,r^{t}_{N},\cdots,r^{t}_{k-1}|\Pi|r^{t'}_{k+1},\cdots,r^{t'}_{N},\cdots,r^{t'}_{k-1}\rangle\\
=&0.
\end{aligned}
\end{equation}
Suppose that $(r^{t}_{k+1},\cdots,r^{t}_{N},\cdots,r^{t}_{k-1})\in\mathcal{G}^{N-1}_{[v,V_{\gamma}]}$ and  $(r^{t'}_{k+1},\cdots,r^{t'}_{N},\cdots,r^{t'}_{k-1})\in\mathcal{G}^{N-1}_{[w,W_{\beta}]}$, where $v\neq w\in \mathcal{O}$, we have $|\textbf{\emph{v}}\rangle=|v^{V_{\gamma}}_{k},r^{t}_{k+1},\cdots,r^{t}_{N},\cdots,r^{t}_{k-1}\rangle\in \mathcal{B}_{v}$,  $|\textbf{\emph{w}}\rangle=|w^{W_{\beta}}_{k},r^{t'}_{k+1},\cdots,r^{t'}_{N},\cdots,r^{t'}_{k-1}\rangle\in \mathcal{B}_{w}$ and $v^{V_{\gamma}}_{k}=w^{W_{\beta}}_{k}$, thus
\begin{equation}\label{eq:w}
\small{
\begin{aligned}
&\langle\textbf{\emph{r}}|E|\textbf{\emph{r}}'\rangle\\
=&\langle r^{t}_{k+1},\cdots,r^{t}_{N},\cdots,r^{t}_{k-1}|\Pi|r^{t'}_{k+1},\cdots,r^{t'}_{N},\cdots,r^{t'}_{k-1}\rangle\\
=&\langle\textbf{\emph{v}}|\mathbb{I}\otimes\Pi|\textbf{\emph{w}}\rangle\\
=&0,
\end{aligned}
}
\end{equation}
the last step is obtained by Eq. (\ref{eq:rs}), so Lemma \ref{le-trivial} is satisfied.$\hfill\blacksquare$

\begin{theorem}\label{th-2}
Consider the set $\mathcal{S}=\bigcup_{r\in Q}\mathcal{S}_{r}$, where $\mathcal{S}_{r}$ is given in Eq.\;(\ref{eq:ss}). Suppose $\mathcal{S}$ is symmetric and $\bigcup_{r\in \mathcal{Q}}\bigcup_{\tau\in\mathbb{Z}_{m^{r}}}\mathcal{G}^{N-1}_{[r,\tau]}=\mathbb{Z}_{d_{k+1}}\times\cdots\times\mathbb{Z}_{d_{N}}\times\cdots\times\mathbb{Z}_{d_{k-1}}$.
When $E_{\mathcal{B}_{r}}$ is proportional to the identity operator $\mathbb{I}_{r}$,
$\Pi$ can only be trivial if and only if the set $\bigcup_{r\in Q}\mathcal{G}^{N}_{r}$ (\ref{eq:GNr}) satisfy the following conditions:

(i) For any two different strings $(i_{k+1},\cdots, i_{N},\cdots, i_{k-1})$, $(j_{k+1}, \cdots, j_{N}, \cdots, j_{k-1})\in \mathbb{Z}_{d_{k+1}}\times\cdots\times\mathbb{Z}_{d_{N}}\times\cdots\times\mathbb{Z}_{d_{k-1}}$, at least one element $i_{k}$ in $\mathbb{Z}_{d_{k}}$ such that $(i_{k},i_{k+1},\cdots, i_{N},\cdots,i_{k-1})$, $(i_{k},j_{k+1},\cdots, j_{N},\cdots, j_{k-1})\in\bigcup_{r\in\mathcal{Q}}\mathcal{G}^{N}_{r}$.

(ii) The family of sets $\{\bigcup_{\tau\in \mathbb{Z}_{m^{r}}}\mathcal{G}^{N-1}_{[r,R_{\tau}]}\}_{r\in\mathcal{Q}}$ are connected.
\end{theorem}
$\textit{Proof.}$
$\textit{Sufficient}.$
We first show that $\Pi$ can only be trivial. Since $E_{\mathcal{B}_{r}}\propto \mathbb{I}_{\mathcal{B}_{r}}$, for any two different states $|\textbf{\emph{i}} \rangle$, $|\textbf{\emph{i}}' \rangle\in \mathcal{B}_{r}$, we get $\langle\textbf{\emph{i}}|E|\textbf{\emph{i}}'\rangle=\langle\textbf{\emph{i}}'|E|\textbf{\emph{i}}\rangle=0$.
It follows from Lemma \ref{le-zeros} that the states $|\textbf{\emph{i}} \rangle\in\mathcal{B}_{r}$ and $|\textbf{\emph{j}} \rangle\in \mathcal{B}_{v}$ $(r\neq v\in\mathcal{Q})$ satisfy $\langle\textbf{\emph{i}}|E|\textbf{\emph{j}}\rangle=\langle\textbf{\emph{j}}|E|\textbf{\emph{i}}\rangle=0$.

So, for any two different states $|\textbf{\emph{i}}\rangle$, $|\textbf{\emph{j}}\rangle \in\bigcup_{r\in\mathcal{Q}}\mathcal{B}_{r}$, we have
\begin{equation}\label{eq:12}
\begin{aligned}
\langle\textbf{\emph{i}}|E|\textbf{\emph{j}}\rangle=\langle\textbf{\emph{j}}|E|\textbf{\emph{i}}\rangle=0.
\end{aligned}
\end{equation}
By condition (i), substituting $|\textbf{\emph{i}}\rangle=|i_{k},i_{k+1},\cdots, i_{N},\cdots,i_{k-1}\rangle$ and $|\textbf{\emph{j}}\rangle=|i_{k},j_{k+1},\cdots, j_{N},\cdots,j_{k-1}\rangle$ into Eq. (\ref{eq:12}), one gets
\begin{equation}\label{eq:14}
\begin{aligned}
&\langle i_{k},i_{k+1},\cdots, i_{N},\cdots,i_{k-1}|E|i_{k},j_{k+1},\cdots, j_{N},\cdots,j_{k-1}\rangle\\
=&\langle i_{k+1},\cdots, i_{N},\cdots,i_{k-1}|\Pi|j_{k+1},\cdots, j_{N},\cdots,j_{k-1}\rangle\\
=&0,
\end{aligned}
\end{equation}
thus the off-diagonal elements of $\Pi$ are all zeros.

Furthermore, as $E_{\mathcal{B}_{r}}\propto \mathbb{I}_{\mathcal{B}_{r}}$, for two different vectors $|\textbf{\emph{i}}\rangle=|i_{k},i_{k+1},\cdots,i_{N},\cdots,i_{k-1}\rangle$ and $|\textbf{\emph{i}}'\rangle=|i'_{k},i'_{k+1},\cdots,i'_{N},\cdots,i                                                            '_{k-1}\rangle$ belong to $\mathcal{B}_{r}$, we have
\begin{equation}\label{eq:15}
\begin{aligned}
\langle\textbf{\emph{i}}|E|\textbf{\emph{i}}\rangle=\langle\textbf{\emph{i}}'|E|\textbf{\emph{i}}'\rangle.
\end{aligned}
\end{equation}
By condition (i), let $i_{k}=i'_{k}$ in Eq. (\ref{eq:15}), we get
\begin{equation}\label{eq:16}
\begin{aligned}
&\langle i_{k+1},\cdots,i_{N},\cdots,i_{k-1}|\Pi|i_{k+1},\cdots,i_{N},\cdots,i_{k-1}\rangle\\
=&\langle i'_{k+1},\cdots,i'_{N},\cdots,i'_{k-1}|\Pi|i'_{k+1},\cdots,i'_{N},\cdots,i'_{k-1}\rangle
\end{aligned}
\end{equation}
for any two different strings $(i_{k+1},\cdots,i_{N},\cdots,i_{k-1})$, $(i'_{k+1},\cdots,i'_{N},\cdots,i'_{k-1})\in\bigcup_{\tau\in \mathbb{Z}_{m^{r}}}\mathcal{G}^{N-1}_{[r,R_{\tau}]}$. Since $\{\bigcup_{\tau\in\mathbb{Z}_{m^{r}}}\mathcal{G}^{N-1}_{[r,\tau]}\}_{r\in \mathcal{Q}}$ is connected, for any different $(N-1)$-tuples $(i_{k+1},\cdots, i_{N},\cdots,i_{k-1})\neq (j_{k+1},\cdots, j_{N},\cdots,j_{k-1})\in\{\bigcup_{\tau\in\mathbb{Z}_{m^{r}}}\mathcal{G}^{N-1}_{[r,\tau]}\}_{r\in \mathcal{Q}}$, we have
\begin{equation}\label{eq:16}
\begin{aligned}
&\langle i_{k+1},\cdots,i_{N},\cdots,i_{k-1}|\Pi|i_{k+1},\cdots,i_{N},\cdots,i_{k-1}\rangle\\
=&\langle j_{k+1},\cdots,j_{N},\cdots,j_{k-1}|\Pi|j_{k+1},\cdots,j_{N},\cdots,j_{k-1}\rangle,
\end{aligned}
\end{equation}
thus the diagonal elements of $\Pi$ are all equal. Accordingly, we get $\Pi\varpropto \mathbb{I}$.

$\textit{Necessity.}$ By Eq. (\ref{eq:ss}), $\mathcal{S}_{r}$ is an orthogonal set spanned by $\mathcal{B}_{r}$, Dim $(\mathrm{span}~\mathcal{S}_{r})=|\mathbb{Z}_{\mathcal{B}_{r}}|$, then
\begin{equation}
\mathrm{span}~\{|\Psi_{r,s}\rangle\}_{s\in\mathbb{Z}_{|\mathcal{B}_{r}|}}=\mathrm{span}~\mathcal{B}_{r}.
\end{equation}
Thus for $|\textbf{\emph{i}}\rangle\in\mathcal{B}_{r}$, it has a linear combination of $\{|\Psi_{r,s}\rangle\}_{s\in\mathbb{Z}_{|\mathcal{B}_{r}|}}$, by Eq. (\ref{eq:rs}),
one obtains
\begin{equation}\label{eq:ij=0}
\begin{aligned}
&\langle \textbf{\emph{i}}|\mathbb{I}\otimes\Pi|\textbf{\emph{j}}\rangle\\
=&\langle i_{k}|j_{k}\rangle\langle i_{k+1},\cdots,i_{k-1}|\Pi|j_{k+1},\cdots,j_{k-1}\rangle\\
=&0
\end{aligned}
\end{equation}
for any different vectors $|\textbf{\emph{i}}\rangle=|i_{k},i_{k+1},\cdots, i_{N},\cdots,i_{k-1}\rangle$, $|\textbf{\emph{j}}\rangle=|j_{k},j_{k+1},\cdots, j_{N},\cdots, j_{k-1}\rangle\in\bigcup_{r\in\mathcal{Q}}\mathcal{B}_{r}$.
Since $\Pi$ is proportional to the identity operator, when $(i_{k+1},\cdots, i_{N},\cdots,i_{k-1})=(j_{k+1},\cdots, j_{N},\cdots, j_{k-1})$  we get $\langle i_{k+1},\cdots,i_{k-1}|\Pi|j_{k+1},\cdots,j_{k-1}\rangle\neq0$, then $\langle i_{k}|j_{k}\rangle=0$, that is $i_{k}\neq j_{k}$. When $(i_{k+1},\cdots, i_{N},\cdots,i_{k-1})\neq(j_{k+1},\cdots, j_{N},\cdots, j_{k-1})$  there is $\langle i_{k+1},\cdots,i_{k-1}|\Pi|j_{k+1},\cdots,j_{k-1}\rangle=0$,
then $\langle i_{k}|j_{k}\rangle$ equals $0$ or $1$. In this case, we will show that there must have $\langle i_{k}|j_{k}\rangle=1$ by contradiction.
Assuming that $\langle i_{k}|j_{k}\rangle$ is only equal to $0$, we get $i_{k}\neq j_{k}$. By the symmetry of $\mathcal{S}$, it follows that the vectors $|i_{k+1},\cdots, i_{N},\cdots,i_{k-1},i_{k}\rangle$ and $|j_{k+1},\cdots, j_{N},\cdots, j_{k-1},j_{k}\rangle$  belong to set $\bigcup_{r\in\mathcal{Q}}\mathcal{B}_{r}$. Substituting these two vectors to Eq.\;(\ref{eq:ij=0}), we get $i_{k+1}\neq j_{k+1}$ by the assumption. By the same token, one can deduce that
\begin{equation*}
\begin{aligned}
\begin{cases}
i_{k+2}\neq j_{k+2},\\
~~~~\vdots\\
i_{N}\neq j_{N},\\
~~~~\vdots\\
i_{k-1}\neq j_{k-1}.
\end{cases}
\end{aligned}
\end{equation*}
It follows that $\bigcup_{r\in \mathcal{Q}}\bigcup_{\tau\in\mathbb{Z}_{m^{r}}}\mathcal{G}^{N-1}_{[r,\tau]}\neq\mathbb{Z}_{d_{k+1}}\times\cdots\times\mathbb{Z}_{d_{N}}\times\cdots\times\mathbb{Z}_{d_{k-1}}$, this leads to a contradiction.

On the other hand, as $E_{\mathcal{B}_{r}}\propto \mathbb{I}_{\mathcal{B}_{r}}$, for any different vectors $|\textbf{\emph{i}}\rangle=|i_{k},i_{k+1},\cdots, i_{N},\cdots,i_{k-1}\rangle,~|\textbf{\emph{i}}'\rangle=|i'_{k},i'_{k+1},\cdots, i'_{N},\cdots,i'_{k-1}\rangle\in \mathcal{B}_{r}$, we have
\begin{equation}
\begin{aligned}
\langle\textbf{\emph{i}}|E|\textbf{\emph{i}}\rangle=\langle\textbf{\emph{i}}'|E|\textbf{\emph{i}}'\rangle, \end{aligned}
\end{equation}
then
\begin{equation}
\begin{aligned}
&\langle i_{k+1},\cdots, i_{N},\cdots,i_{k-1}|\Pi|i_{k+1},\cdots, i_{N},\cdots,i_{k-1}\rangle\\
=&\langle i'_{k+1},\cdots, i'_{N},\cdots,i'_{k-1}|\Pi|i'_{k+1},\cdots, i'_{N},\cdots,i'_{k-1}\rangle.
\end{aligned}
\end{equation}
Since the diagonal elements of $\Pi$ are all equal, one can deduce that $\{\bigcup_{\tau\in\mathbb{Z}_{m^{r}}}\mathcal{G}^{N-1}_{[r,R_{\tau}]}\}_{r\in\mathcal{Q}}$ is connected. $\hfill\blacksquare$

Theorem \ref{th-2} provides a sufficient and necessary condition for $\mathcal{S}=\bigcup_{r\in Q}\mathcal{S}_{r}$ (\ref{eq:ss}) which has the property of strongest nonlocality. Using the condition (i) of Theorem \ref{th-2}, in $(\mathbb{C}^{3})^{\otimes N}$, we deduce the smallest size of $\mathcal{S}$,
what we need to do is to discuss the smallest size of $\bigcup_{r\in\mathcal{Q}}\mathcal{G}_{r}^{N}$. Let $\mathcal{C}(\mathbb{Z}^{N-1}_{3})$ be the set that satisfies the condition (i) of Theorem \ref{th-2}, $\mathcal{S}\mathcal{C}(\mathbb{Z}^{N-1}_{3})$ be all cyclic symmetric elements of $\mathcal{C}(\mathbb{Z}^{N-1}_{3})$, obviously, $\mathcal{C}(\mathbb{Z}^{N-1}_{3})\subset\mathcal{S}\mathcal{C}(\mathbb{Z}^{N-1}_{3})$.
\begin{theorem}
For the set $\mathcal{S}=\bigcup_{r\in Q}\mathcal{S}_{r}$ (\ref{eq:ss}) in $(\mathbb{C}^{3})^{\otimes N}~(N\geq3)$, if $\mathcal{S}$ is symmetric and any orthogonality-preserving POVM on $\bar{A}_{k}$ can only be trivial, then $\mathcal{S}$ has the strongest nonlocality, the smallest size of this set is $2\times3^{N-1}$.
\end{theorem}
$\textit{Proof.}$
If set $\bigcup_{r\in\mathcal{Q}}\mathcal{G}_{r}^{N}$ contains  $\mathcal{C}(\mathbb{Z}^{N-1}_{3})$, due to the symmetry, it must contain $\mathcal{S}\mathcal{C}(\mathbb{Z}^{N-1}_{3})$. In addition to the case that $\mathcal{C}(\mathbb{Z}^{N-1}_{3})=\mathbb{Z}^{N}_{3}$, four cases of $\mathcal{C}(\mathbb{Z}^{N-1}_{3})$ are as follows.

Case I: For every string $(i_{2},\cdots,i_{N})$ in $\mathbb{Z}^{N-1}_{3}$, if only one element $i_{1}\in\mathbb{Z}_{3}$ such that $(i_{1},i_{2},\cdots,i_{N})\in \mathcal{C}(\mathbb{Z}^{N-1}_{3})$, we get $\mathcal{C}(\mathbb{Z}^{N-1}_{3})=~\{i_{1}\}\times\mathbb{Z}^{N-1}_{3}$ and $|\mathcal{C}(\mathbb{Z}^{N-1}_{3})|=3^{N-1}$. One can check that $\mathcal{S}\mathcal{C}(\mathbb{Z}^{N-1}_{3})$ is the collection of $N$-tuples that at least one component is $i_{1}$, thus  $\mathcal{S}\mathcal{C}(\mathbb{Z}^{N-1}_{3})$ has size $3^{N}-2^{N}$. Therefore, the size of $\bigcup_{r\in\mathcal{Q}}\mathcal{G}_{r}^{N}$ is not less than $3^{N}-2^{N}$.

Case II: For every string $(i_{2},\cdots,i_{N})$ in $\mathbb{Z}^{N-1}_{3}$, there are exactly two different elements $i_{1},~i'_{1}\in\mathbb{Z}_{3}$ such that $(i_{1},i_{2},\cdots,i_{N}),~(i'_{1},i_{2},\cdots,i_{N})\in \mathcal{C}(\mathbb{Z}^{N-1}_{3})$. We get $\mathcal{C}(\mathbb{Z}^{N-1}_{3})=~\{i_{1},i'_{1}\}\times\mathbb{Z}^{N-1}_{3}$  and has size $2\times3^{N-1}$. Since $\mathcal{C}(\mathbb{Z}^{N-1}_{3})\subset\mathcal{S}\mathcal{C}(\mathbb{Z}^{N-1}_{3})$, then the size of $\mathcal{S}\mathcal{C}(\mathbb{Z}^{N-1}_{3})$ is not less than $2\times3^{N-1}$,
thus the size of $\bigcup_{r\in\mathcal{Q}}\mathcal{G}_{r}^{N}$ cannot be less than $2\times3^{N-1}$.

Let $\mathbb{P}^{N-1}$ is a subset of $\mathbb{Z}^{N-1}_{3}$ and $0<|\mathbb{P}^{N-1}|<3^{N-1}$.

Case III: For every string $(p_{2},\cdots,p_{N})$ in $\mathbb{P}^{N-1}$, there is only one element $p_{1}\in\mathbb{Z}_{3}$ such that $(p_{1},p_{2},\cdots,p_{N})\in\mathcal{C}(\mathbb{Z}^{N-1}_{3})$. In addition, for every string $(q_{2},\cdots,q_{N})$ in $\mathbb{Z}^{N-1}_{3}\backslash \mathbb{P}^{N-1}$, there is a subset $\mathbb{Q}$ of $\mathbb{Z}_{3}~(|\mathbb{Q}|>1)$ such that $\mathbb{Q}\times(q_{2},\cdots,q_{N})\subset \mathcal{C}(\mathbb{Z}^{N-1}_{3})$. By  condition (i) of Theorem \ref{th-2}, we get $p_{1}\in\mathbb{Q}$, thus $\{p_{1}\}\times\mathbb{Z}^{N-1}_{3}\subset \mathcal{C}(\mathbb{Z}^{N-1}_{3})$. As stated in Case I, the size of $\bigcup_{r\in\mathcal{Q}}\mathcal{G}_{r}^{N}$ is not less than $3^{N}-2^{N}$.

Case IV: For every string $(p_{2},\cdots,p_{N})$ in $\mathbb{P}^{N-1}$, there are two different elements $p_{1},~p'_{1}\in\mathbb{Z}_{3}$ such that $(p_{1},p_{2},\cdots,p_{N}),~(p'_{1},p_{2},\cdots,p_{N})\in\mathcal{C}(\mathbb{Z}^{N-1}_{3})$. For the string $(q_{2},\cdots,q_{N})\in \mathbb{Z}^{N-1}_{3}\backslash \mathbb{P}^{N-1}$ and  the subset $\mathbb{Q}$ of $\mathbb{Z}_{3}~(|\mathbb{Q}|>1)$, there is $\mathbb{Q}\times(q_{2},\cdots,q_{N})\subset \mathcal{C}(\mathbb{Z}^{N-1}_{3})$. As $|\mathbb{Q}|=2$ is more likely than $|\mathbb{Q}|=3$ to get the smallest size of $\bigcup_{r\in\mathcal{Q}}\mathcal{G}_{r}^{N}$, we obtain $|\mathcal{C}(\mathbb{Z}^{N-1}_{3})|=2\times3^{N-1}$, thus the size of $\mathcal{C}\mathcal{S}(\mathbb{Z}^{N-1}_{3})$ is not less than $2\times3^{N-1}$, then the size of $\bigcup_{r\in\mathcal{Q}}\mathcal{G}_{r}^{N}$ cannot be less than $2\times3^{N-1}$.

Actually, in system $(\mathbb{C}^{3})^{\otimes N}$, the set in Ref.\;\cite{Quantum.7.1101} corresponds to Case I and has size $3^{N}-2^{N}$.
Obviously, $2\times3^{N-1}$ is fewer to $3^{N}-2^{N}$, meanwhile, the strongest nonlocal genuinely entangled set in Ref.\;\cite{PRA.109.022220} satisfies Theorem \ref{th-2} and contains $2\times3^{N-1}$ states. Consequently, in $(\mathbb{C}^{3})^{\otimes N}$ , the minimum size of the set $\mathcal{S}=\bigcup_{r\in Q}\mathcal{S}_{r}$ is $2\times3^{N-1}$.$\hfill\blacksquare$

\section{ ORTHOGONAL GENUINELY ENTANGLED SETS WITH STRONGEST NONLOCALITY IN $(\mathbb{C}^{d})^{\otimes N}$}\label{Q3}
In this section, we generalize the strongest nonlocal genuinely entangled set in Ref. \cite{PRA.109.022220} to $(\mathbb{C}^{d})^{\otimes N}~(d\geq4)$ and show that there exist smaller strongest nonlocal OGESs in high-dimensional multipartite systems.
\subsection{An ORTHOGONAL GENUINELY ENTANGLED BASE in $(\mathbb{C}^{d})^{\otimes N}$}
For each element $i\in\mathbb{Z}_{d}$, let
\begin{equation}\label{eq:G1}
\mathcal{G}^{1}_{i}:=\{i\},
\end{equation}
then we construct $d$ subsets of $\mathbb{Z}^{N}_{d}$ for $N\geq2$,
\begin{equation}\label{eq:GN}
\mathcal{G}^{N}_{i}:=\bigcup^{d-1}_{j=0}\left(\{i\oplus_{d}(d-j)\}\times\mathcal{G}^{N-1}_{j}\right),
\end{equation}
where $i\in\mathbb{Z}_{d}$, $i\oplus_{d}t=(i+t)$ mod $d$.
\begin{proposition}\label{pr1}
The sets in Eq.\;(\ref{eq:GN}) are pairwise disjoint and the union of all is $\mathbb{Z}^{N}_{d}$, i.e.,
\begin{equation*}
\bigcup^{d-1}_{i=0}\mathcal{G}^{N}_{i}=\mathbb{Z}^{N}_{d}~\mathrm{and}~
\mathcal{G}^{N}_{i}\cap\mathcal{G}^{N}_{j}=\emptyset, ~\mathrm{where}~i\neq j\in\mathbb{Z}_{d}.
\end{equation*}
\end{proposition}
The detailed proof is shown in Appendix \ref{A}.

\begin{proposition}\label{pr2}
The set $\mathcal{G}^{N}_{i}$ in Eq. (\ref{eq:GN}) is invariant under arbitrary permutation of the positions of the $N$ components.
\end{proposition}
The detailed proof is shown in Appendix \ref{B}.

Let $\mathcal{H}:=(\mathbb{C}^{d})^{\otimes N}$,
\begin{equation}\label{eq:Si}
\begin{aligned}
&\mathcal{S}_{i}:=\{|\Psi_{i,k}\rangle \in \mathcal{H}~\big|~k \in \mathbb{Z}_{s_{i}}, |\Psi_{i,k}\rangle:=\sum\limits_{\textbf{\emph{j}} \in {\mathcal{G}^{N}_{i}}}\omega^{k f_{i}(\textbf{\emph{j}})}_{s_{i}}|\textbf{\emph{j}}\rangle\}.\\
\end{aligned}
\end{equation}
 Here $ f_{i}:\mathcal{G}^{N}_{i}\longrightarrow \mathbb{Z}_{s_{i}}$ is any fixed bijection and $\omega_{s_{i}}:=\mathrm{e}^{\frac{2\pi\sqrt{-1}}{s_{i}}}$, $s_{i}=|\mathcal{G}^{N}_{i}|~(i\in \mathbb{Z}_{d})$.
\begin{proposition}
The set $\bigcup^{d-1}_{i=0}\mathcal{S}_{i}$ of states given by Eq. (\ref{eq:Si}) is an orthogonal genuinely entangled base in $(\mathbb{C}^{d})^{\otimes N}$.
\end{proposition}
$\textit{Proof.}$
By Proposition \ref{pr1}, it follows that $\bigcup^{d-1}_{i=0}\mathcal{S}_{i}$ is a base in system $(\mathbb{C}^{d})^{\otimes N}$. The proof of genuine entanglement of $|\Psi_{i,k}\rangle$ is similar to Ref. [\cite{PRA.109.022220},~Theorem~1]. $\hfill\blacksquare$
\subsection{STRONGEST NONLOCAL OGESs in $(\mathbb{C}^{d})^{\otimes N}~(d\geq4)$}
Based on the above result, we present the strongest nonlocal genuinely entangled sets in system $(\mathbb{C}^{d})^{\otimes N}$.
We first consider which sets are sufficient to satisfy the condition (i) of Theorem \ref{th-2}. From Eq.\;(\ref{eq:GNandGN-1}),
the relationship between $\{\mathcal{G}^{N}_{i}\}_{i\in\mathbb{Z}_{d}}$ and $\{\mathcal{G}^{N-1}_{j}\}_{j\in\mathbb{Z}_{d}}$ is described by $\mathcal{M}_{d}=\{m_{i,j}\}_{i,j\in \mathbb{Z}_{d}}$ (\ref{eq:Md}). For simplicity, we label row $i$ by set $\mathcal{G}^{N}_{i}$ and column $j$ by set $\mathcal{G}^{N-1}_{j}$.
\begin{spacing}{1.3}
\begin{align*}
\mathcal{M}_{d}=
\begin{matrix}
&\mathcal{G}^{N-1}_{0} & \mathcal{G}^{N-1}_{1} & \mathcal{G}^{N-1}_{2}  & \cdots & \mathcal{G}^{N-1}_{d-2} &\mathcal{G}^{N-1}_{d-1}\\
\mathcal{G}^{N}_{0} \ldelim[{6}{0cm} &0  & d-1 & d-2 & \cdots & 2 & 1 &\rdelim]{6}{0.1cm} \\
\mathcal{G}^{N}_{1}  ~&1  & 0 & d-1 & \cdots & 3 & 2\\
\mathcal{G}^{N}_{2}  ~&2  & 1 & 0 & \cdots & 4 & 3\\
\vdots ~&  \vdots & \vdots & \vdots&\ddots&\vdots&~\vdots\\
\mathcal{G}^{N}_{d-2} ~& d-2 & d-3 & d-4 & \cdots & 0 & d-1\\
\mathcal{G}^{N}_{d-1} ~&  d-1  & d-2 & d-3 & \cdots & 1 & 0
\end{matrix}.
\end{align*}
\end{spacing}
The distance of any two  rows is defined as
\begin{equation*}
r~(\mathcal{G}^{N}_{i_{1}}, \mathcal{G}^{N}_{i_{2}})=\mathrm{min}\{|i_{1}-i_{2}|,~d-|i_{1}-i_{2}|\},
\end{equation*}
where $i_{1}\neq i_{2}\in \mathbb{Z}_{d}$, then one gets
\begin{equation*}
 r~(\mathcal{G}^{N}_{i_{1}}, \mathcal{G}^{N}_{i_{2}})~\in
 \begin{aligned}
 \left[1,\lfloor\frac{d}{2}\rfloor\right],
 \end{aligned}
\end{equation*}
with $\lfloor.\rfloor$ denoting the floor function.
The distance of any two columns is denoted as $c~(\mathcal{G}^{N-1}_{j_{1}}, \mathcal{G}^{N-1}_{j_{2}})$ and has the same definition as $r~(\mathcal{G}^{N}_{i_{1}}, \mathcal{G}^{N}_{i_{2}})$. Then we get the following observations.

\begin{Ob}\label{ob1}
For any $2\times 2$ submatrix $P_{2\times 2}$ of $\mathcal{M}_{d}$,
\begin{spacing}{1.3}
\begin{equation*}
P_{2\times 2}=\bbordermatrix{
                          &\mathcal{G}^{N-1}_{j_{1}} & \mathcal{G}^{N-1}_{j_{2}}\cr
\mathcal{G}^{N}_{i_{1}} & m_{i_{1},j_{1}}          & m_{i_{1},j_{2}}\cr
\mathcal{G}^{N}_{i_{2}} &m_{i_{2},j_{1}}           & m_{i_{2},j_{2}}
},
\end{equation*}
\end{spacing}
\noindent if $r ~(\mathcal{G}^{N}_{i_{1}}, \mathcal{G}^{N}_{i_{2}})$ = $c~(\mathcal{G}^{N-1}_{j_{1}}, \mathcal{G}^{N-1}_{j_{2}})$ we get $m_{i_{1},j_{1}}= m_{i_{2},j_{2}}$ or $ m_{i_{1}, j_{2}}=m_{i_{2},j_{1}}$, i.e., there are the same elements in both columns.
\end{Ob}
\begin{Ob}\label{ob2}
For any two different columns, since the union of all distances is $\bigcup_{j_{1}\neq j_{2}\in \mathbb{Z}_{d}}\{c~(\mathcal{G}^{N-1}_{j_{1}}, \mathcal{G}^{N-1}_{j_{2}})\}=\left[1,\lfloor\frac{d}{2}\rfloor\right]$,
if we can find rows $\{\mathcal{G}^{N}_{i}\}_{i\in T\subset \mathbb{Z}_{d}}$
such that $\bigcup_{i_{1}\neq i_{2}\in T}\{r(\mathcal{G}^{N}_{i_{1}}, \mathcal{G}^{N}_{i_{2}})\}= \left[1,\lfloor\frac{d}{2}\rfloor\right]$, it follows from Observation \ref{ob1} that condition (i) of Theorem \ref{th-2} is satisfied.
\end{Ob}

\begin{proposition}\label{prrows}
 The rows $\{\mathcal{G}^{N}_{t}\}_{t\in T_{1}\cup T_{2}}$, where $T_{1}=\{t_{1}|0<t_{1}<\lfloor\frac{d}{2}\rfloor, ~t_{1}~\mathrm{is}~\mathrm{odd}\}$ and $T_{2}=\{0,\lfloor\frac{d}{2}\rfloor\}$, satisfy the condition (i) of Theorem \ref{th-2}.
\end{proposition}
\textit{$Proof.$}
When $\lfloor\frac{d}{2}\rfloor$ is odd, we have
\begin{equation*}
\begin{aligned}
\begin{cases}
\bigcup_{t\in T_{1}}\{r(\mathcal{G}^{N}_{0},\mathcal{G}^{N}_{t})\}=\{1,3,\cdots,\lfloor\frac{d}{2}\rfloor-2\},\\
\bigcup_{t\in T_{1}\backslash\{1\}}\{r(\mathcal{G}^{N}_{1},\mathcal{G}^{N}_{t})\}=\{2,4,\cdots,\lfloor\frac{d}{2}\rfloor-3\},\\
r(\mathcal{G}^{N}_{1},\mathcal{G}^{N}_{\lfloor\frac{d}{2}\rfloor})=\lfloor\frac{d}{2}\rfloor-1,\\
r(\mathcal{G}^{N}_{0},\mathcal{G}^{N}_{\lfloor\frac{d}{2}\rfloor})=\lfloor\frac{d}{2}\rfloor.\\
\end{cases}
\end{aligned}
\end{equation*}
When $\lfloor\frac{d}{2}\rfloor$ is even, we have
\begin{equation*}
\begin{aligned}
\begin{cases}
\bigcup_{t\in T_{1}}\{r(\mathcal{G}^{N}_{0},\mathcal{G}^{N}_{t})\}=\{1,3,\cdots,\lfloor\frac{d}{2}\rfloor-1\},\\
\bigcup_{t\in T_{1}\backslash\{1\}}\{r(\mathcal{G}^{N}_{1},\mathcal{G}^{N}_{t})\}=\{2,4,\cdots,\lfloor\frac{d}{2}\rfloor-2\},\\ r(\mathcal{G}^{N}_{0},\mathcal{G}^{N}_{\lfloor\frac{d}{2}\rfloor})=\lfloor\frac{d}{2}\rfloor.
\end{cases}
\end{aligned}
\end{equation*}
By Observation \ref{ob2}, the condition (i) of Theorem \ref{th-2} holds. $\hfill\blacksquare$

To complete the construction, we need to get the distribution of element $(\xi)^{\times N}(\xi\in\mathbb{Z}_{d})$ in set $\mathcal{G}^{N}_{j}$, we first recall several knowledge of cyclic group.

$Cyclic~group$. A group $G$ is said to be cyclic if each element can be generated by $g\in G$, the element $g$ is called  a generator of $G$ and denoted $G=\langle g\rangle$.

\begin{lemma}\label{lem-group}
Let $G=\langle g\rangle$ is a cyclic group of order $n$, then the order of element $g^{s}$ is $\frac{n}{\mathrm{gcd}(s,n)}$, where $\mathrm{gcd}(s,n)$ is the greatest common divisor of $s$ and $n$.
\end{lemma}
\begin{theorem}\label{th-element}
Every set in $\{\mathcal{G}^{N}_{t}\}_{t\in T_{1}\cup T_{2}}$ contains $N$-tuples of the form $(\xi')^{\times N}$ $(\xi'\in \mathbb{Z}_{d}\backslash\{0\})$.
\end{theorem}
$\textit{Proof.}$
By Eq.\;(\ref{eq:ZNi}), we first obtain the distribution of element $(\xi)^{\times N}(\xi\in\mathbb{Z}_{d})$ in $\mathcal{G}^{N}_{j}~(j\in\mathbb{Z}_{d})$. Here we classify $N$ by congruence of modulo $d$, let $a\equiv N~(\mathrm{mod}~d)$, denoted as $[N]=[a],~a\in \mathbb{Z}_{d}$. Then we find the relationship of component $\xi$, classification $a$ and the subscript $j$ of the set $\mathcal{G}^{N}_{j}$ is
\begin{equation}\label{eq:ja}
j=a\oplus_{d}a\oplus_{d}\cdots\oplus_{d}a,
\end{equation}
where $a$ is repeated $\xi$ times. The detailed distribution is shown in Table \ref{tab:distri}.
To complete the proof we need to show that each row in Table \ref{tab:distri} from column $(1)^{\times N}$ to column $(d-1)^{\times N}$ contains the set $\mathcal{G}^{N}_{j_{0}}$, and the subscript $j_{0}$ belongs to set $T_{1}\cup T_{2}$.
\begin{table}\centering
\caption{The distribution of $(\xi)^{\times N}$, where $\xi\in \mathbb{Z}_{d}$.}
\renewcommand\arraystretch{1.2}
\tabcolsep=0.01cm
\begin{tabular}{l c  c  c  c  c  c}
 \hline
 \hline
System $N$& $(0)^{\times N}$ & $(1)^{\times N}$ & $(2)^{\times N}$  & $\cdots$   &$(d-2)^{\times N}$ &$(d-1)^{\times N}$\\
 \hline
$[N]=[0]$ & $\mathcal{G}^{N}_{0}$  & $\mathcal{G}^{N}_{0}$ & $\mathcal{G}^{N}_{0}$  & $\cdots$ & $\mathcal{G}^{N}_{0}$ & $\mathcal{G}^{N}_{0}$\\

$[N]=[1]$ & $\mathcal{G}^{N}_{0}$  & $\mathcal{G}^{N}_{1}$ & $\mathcal{G}^{N}_{2}$ & $\cdots$ & $\mathcal{G}^{N}_{d-2}$ & $\mathcal{G}^{N}_{d-1}$\\

$[N]=[2]$ & $\mathcal{G}^{N}_{0}$  & $\mathcal{G}^{N}_{2}$ & $\cdots$ &   &  &\\

~~~~~~~~\vdots &  & \\

$[N]=[d-2]$ & $\mathcal{G}^{N}_{0}$  & $\mathcal{G}^{N}_{d-2}$ & $\cdots$ &   & &\\

$[N]=[d-1]$ & $\mathcal{G}^{N}_{0}$  & $\mathcal{G}^{N}_{d-1}$ &$\mathcal{G}^{N}_{d-2}$ & $\cdots$ & $\mathcal{G}^{N}_{2}$ & $\mathcal{G}^{N}_{1}$\\

\hline
\hline

\end{tabular}\label{tab:distri}
\end{table}~\

From Eq.\;(\ref{eq:ja}), we get that $\{[j]\}_{j}=\langle [a]\rangle$ is a cyclic subgroup of group $\{[0],[1],\cdots,[d-1]\}$.
Using Lemma \ref{lem-group}, we obtain the order of $\langle [a]\rangle$ is $k=\frac{d}{\mathrm{gcd}(s,d)}$, where $[a]=s[1]~(0\leq s\leq d-1)$. We will discuss rows $[N]=[a]$ in the following three cases.

When $a=0$, we get $j=0$ by Eq. (\ref{eq:ja}), it follows that the sets in row $[N]=[0]$ are all the same and are the $\mathcal{G}^{N}_{0}$.

When $a$ and $d$ are coprime, the order of $\langle [a]\rangle$ is $d$, we get $\{[j]\}_{j}=\{[0],[1],\cdots,[d-1]\}$, which has an element $\lfloor\frac{d}{2}\rfloor$ belongs to $T_{2}$.

When $a$ is a factor of $d$ and $1<a<d$, we get the order $k$ of $\langle [a]\rangle$ satisfies $k|d$ and $1<k<d$. Because there are $d$ columns through column $(0)^{\times N}$ to column $(d-1)^{\times N}$, the subscript $0\in T_{2}$ will appear at least twice. The proof is completed. $\hfill\blacksquare$

Now we give the construction of genuinely entangled states in $(\mathbb{C}^{d})^{\times  N}$, where $N$ is divided into three cases as mentioned above.
\begin{Co}
Case I, $[N]=[0]$. Let
\begin{equation}\label{eq:GN[0]w}
\begin{aligned}
&\widetilde{\mathcal{G}^{N}_{0}}=\mathcal{G}^{N}_{0}\backslash\{(0)^{\times N},(\xi')^{\times N}\},\\
&\{\widetilde{\mathcal{G}^{N}_{t}}\}_{t\in T_{1}}=\{\mathcal{G}^{N}_{t}\}_{t\in T_{1}},\\
&\widetilde{\mathcal{G}^{N}_{\lfloor\frac{d}{2}\rfloor}}=\mathcal{G}^{N}_{\lfloor\frac{d}{2}\rfloor},\\
&\widetilde{\mathcal{G}^{N}_{\lfloor\frac{d}{2}\rfloor+1}}=\{0\}\times\{(0)^{\times (N-1)}\}\bigcup \{\xi'\}\times\{(\xi')^{\times (N-1)}\},
\end{aligned}
\end{equation}
where~$\xi'\in \mathbb{Z}_{d}\backslash\{0\}$~and~satisfies~$[0]=\xi'[0]$.

Case II, $[N]=[a]$, where $\mathrm{gcd}(a,d)=1$. Let
\begin{equation}\label{eq:GN[a]w}
\begin{aligned}
&\widetilde{\mathcal{G}^{N}_{0}}=\mathcal{G}^{N}_{0}\backslash\{(0)^{\times N}\},\\
&\{\widetilde{\mathcal{G}^{N}_{t}}\}_{t\in T_{1}}=\{\mathcal{G}^{N}_{t}\}_{t\in T_{1}},\\
&\widetilde{\mathcal{G}^{N}_{\lfloor\frac{d}{2}\rfloor}}=\mathcal{G}^{N}_{\lfloor\frac{d}{2}\rfloor}\backslash\{(\xi')^{\times N}\},\\
&\widetilde{\mathcal{G}^{N}_{\lfloor\frac{d}{2}\rfloor+1}}=\{0\}\times\{(0)^{\times (N-1)}\}\bigcup \{\xi'\}\times\{(\xi')^{\times (N-1)}\},
\end{aligned}
\end{equation}
where~$\xi'\in \mathbb{Z}_{d}\backslash\{0\}$~and~satisfies~$[\lfloor\frac{d}{2}\rfloor]=\xi'[a]$.

Case III, $[N]=[a]$, where $a$ divides $d$, $a\neq1,d$. Let
\begin{equation}\label{eq:GN[a|d]w}
\begin{aligned}
&\widetilde{\mathcal{G}^{N}_{0}}=\mathcal{G}^{N}_{0}\backslash\{(0)^{\times N},(\xi')^{\times N}\},\\
&\{\widetilde{\mathcal{G}^{N}_{t}}\}_{t\in T_{1}}=\{\mathcal{G}^{N}_{t}\}_{t\in T_{1}},\\
&\widetilde{\mathcal{G}^{N}_{\lfloor\frac{d}{2}\rfloor}}=\mathcal{G}^{N}_{\lfloor\frac{d}{2}\rfloor},\\
&\widetilde{\mathcal{G}^{N}_{\lfloor\frac{d}{2}\rfloor+1}}=\{0\}\times\{(0)^{\times (N-1)}\}\bigcup \{\xi'\}\times\{(\xi')^{\times (N-1)}\},
\end{aligned}
\end{equation}
where~$\xi'\in \mathbb{Z}_{d}\backslash\{0\}$~and~satisfies~$[0]=\xi'[a]$.

In each case, let $T_{3}=T_{1}\cup T_{2}\cup\{\lfloor\frac{d}{2}\rfloor+1\}$, $\widetilde{s_{i}}$ be the cardinality of the set $\widetilde{\mathcal{G}^{N}_{i}}$, $i\in T_{3}$, denote
\begin{equation}
\begin{aligned}
&\widetilde{\mathcal{S}_{i}}:=\{|\widetilde{\Psi}_{i,k}\rangle \in \mathcal{H}~\big|~k \in \mathbb{Z}_{\widetilde{s_{i}}}, |\widetilde{\Psi}_{i,k}\rangle:=\sum\limits_{\textbf{\emph{j}}\in {\widetilde{\mathcal{G}^{N}_{i}}}}\omega^{k f_{i}(\textbf{\emph{j}})}_{\widetilde{s_{i}}}|\textbf{\emph{j}}\rangle\}.
\end{aligned}\label{eq:case1Si}
\end{equation}
Here $f_{i}:\widetilde{\mathcal{G}^{N}_{i}}\longrightarrow \mathbb{Z}_{\widetilde{s_{i}}}$ is any fixed bijection and $\omega_{\widetilde{s_{i}}}:=\mathrm{e}^{\frac{2\pi\sqrt{-1}}{\widetilde{s_{i}}}}$.
\end{Co}

Even if we have moved $(0)^{\times N}$ and $(\xi')^{\times N}$, each set in $\{\widetilde{\mathcal{G}^{N}_{t}}\}_{t\in T_{3}}$ is still permutation invariance. Moreover, it follows from the proof of Ref. [\cite{PRA.109.022220},~Theorem~1] that the states in Eq. (\ref{eq:case1Si}) are still genuinely entangled.

\begin{theorem}\label{th-snoges}
In $(\mathbb{C}^{d})^{\times N}$, the states $\widetilde{\mathcal{S}}=\bigcup_{i\in T_{3}}\widetilde{\mathcal{S}_{i}}$ given by Eq. (\ref{eq:case1Si}) possess the strongest nonlocality, and
\begin{equation*}
\begin{aligned}
|\widetilde{\mathcal{S}}|=\begin{cases}
(\frac{\lfloor\frac{d}{2}\rfloor+1}{2} +1)\times d^{(N-1)}, \lfloor\frac{d}{2}\rfloor~\mathrm{is~odd},\\
(\frac{\lfloor\frac{d}{2}\rfloor}{2}+2)\times d^{(N-1)}, \lfloor\frac{d}{2}\rfloor~\mathrm{is~even}.\\
\end{cases}
\end{aligned}
\end{equation*}

\end{theorem}
$\textit{Proof.}$ To prove the strongest nonlocality, we need to show that the $N$-tuples in Eqs.\;(\ref{eq:GN[0]w}--\ref{eq:GN[a|d]w}) satisfy Theorem \ref{th-1} and Theorem \ref{th-2}. Here we give a detailed proof for Case I, the remainder of the argument is analogous to Case I.

For Theorem\;\ref{th-1}, we first consider set $\widetilde{\mathcal{G}^{N}_{\lfloor\frac{d}{2}\rfloor+1}}$, since $\mid \{(0)^{\times (N-1)}\}\mid=1$, one gets that $\widetilde{\mathcal{G}^{N}_{\lfloor\frac{d}{2}\rfloor+1}}$ satisfies the condition (i). Then for any $t\in T_{1}$, $\widetilde{\mathcal{G}^{N}_{t}}$ has a subset $\{t\}\times\mathcal{G}^{N-1}_{0}$ by Eq.\;(\ref{eq:ZNi}), we obatin $\mathcal{G}^{N-1}_{0}=(\mathcal{G}^{N-1}_{0}\backslash\{(0)^{\times (N-1)}\}) \bigcup  \{(0)^{\times (N-1)}\}$ and $|\mathcal{G}^{N-1}_{0}\bigcap\{(0)^{\times (N-1)}\}|=1$, where the first term $\mathcal{G}^{N-1}_{0}\backslash\{(0)^{\times (N-1)}\}$ is taken from the subset $\{0\}\times(\mathcal{G}^{N-1}_{0}\backslash\{(0)^{\times (N-1)}\})$ of $\widetilde{\mathcal{G}^{N}_{0}}$ and the second term $\{(0)^{\times (N-1)}\}$ is taken from the subset $\{0\}\times\{(0)^{\times (N-1)}\}$ of $\widetilde{\mathcal{G}^{N}_{\lfloor\frac{d}{2}\rfloor+1}}$.  Thus $\widetilde{\mathcal{G}^{N}_{t}}$ satisfy the condition (ii). $\widetilde{\mathcal{G}^{N}_{\lfloor\frac{d}{2}\rfloor}}$ satisfies condition (ii) for similar reason to $\widetilde{\mathcal{G}^{N}_{t}}$.

For the set $\widetilde{\mathcal{G}^{N}_{0}}$, $(\{0\}\times\mathcal{G}^{N-1}_{0})\backslash\{(0)^{\times N}\}=\{0\}\times(\mathcal{G}^{N-1}_{0}\backslash\{(0)^{\times (N-1)}\})$ is a subset of it. One can get $\mathcal{G}^{N-1}_{0}\backslash\{(0)^{\times (N-1)}\}\subset\mathcal{G}^{N-1}_{0}$, where $\mathcal{G}^{N-1}_{0}$ on the right is taken from the subset $\{t_{0}\}\times\mathcal{G}^{N-1}_{0}$ of $\widetilde{\mathcal{G}^{N}_{t_{0}}}~(t_{0}\in T_{1})$. Since $\widetilde{\mathcal{G}^{N}_{t_{0}}}$ satisfies the condition (ii), then $\widetilde{\mathcal{G}^{N}_{0}}$ satisfies the condition (iii). $\hfill\blacksquare$

Now we provide an illustrative example of constructing $d=4$, the distribution of element $(\xi)^{\times N}(\xi\in\mathbb{Z}_{4})$ in set $\mathcal{G}^{N}_{j}(j\in\mathbb{Z}_{4})$ is shown in Table \ref{tab:distri2}. For simplicity, we only present the set $\bigcup_{r}\widetilde{\mathcal{G}^{N}_{r}}$ while omitting its corresponding set $\widetilde{\mathcal{S}}=\bigcup_{i\in r}\widetilde{\mathcal{S}_{i}}$.
\begin{table}[h]
\vspace{-1.5em}
\caption{The distribution of $(\xi)^{\times N}$ when $d=4$, where $\xi\in \mathbb{Z}_{4}$.}
\renewcommand\arraystretch{1.2}
\tabcolsep=0.2cm
\begin{tabular}{l c  c  c  c }
 \hline
 \hline
System $N$ & $(0)^{\times N}$ & $(1)^{\times N}$ & $(2)^{\times N}$  &$(3)^{\times N}$ \\
 \hline
$[N]=[0]$ & $\mathcal{G}^{N}_{0}$  & $\mathcal{G}^{N}_{0}$ & $\mathcal{G}^{N}_{0}$  & $\mathcal{G}^{N}_{0}$\\

$[N]=[1]$ & $\mathcal{G}^{N}_{0}$  & $\mathcal{G}^{N}_{1}$ & $\mathcal{G}^{N}_{2}$  & $\mathcal{G}^{N}_{3}$\\

$[N]=[2]$ & $\mathcal{G}^{N}_{0}$  & $\mathcal{G}^{N}_{2}$ & $\mathcal{G}^{N}_{0}$ & $\mathcal{G}^{N}_{2}$\\

$[N]=[3]$ & $\mathcal{G}^{N}_{0}$  & $\mathcal{G}^{N}_{3}$ & $\mathcal{G}^{N}_{2}$ & $\mathcal{G}^{N}_{1}$ \\

\hline
\hline
\end{tabular}\label{tab:distri2}
\end{table}

\vspace{-1.0em}
\textit{Example}~$1$. In $(\mathbb{C}^{4})^{\otimes N}$, the following sets can be the strongest nonlocal OGESs.
\begin{equation*}
\begin{aligned}
&[N]=[0]
\begin{cases}
\widetilde{\mathcal{G}^{N}_{0}}=\mathcal{G}^{N}_{0}\backslash\{(0)^{\times N},~(1)^{\times N}\},\\
\widetilde{\mathcal{G}^{N}_{1}}=\mathcal{G}^{N}_{1},\\
\widetilde{\mathcal{G}^{N}_{2}}=\mathcal{G}^{N}_{2},\\
\widetilde{\mathcal{G}^{N}_{3}}=\{(0)^{\times N},(1)^{\times N}\}.\\
\end{cases}\\
&[N]=[1]
\begin{cases}
\widetilde{\mathcal{G}^{N}_{0}}=\mathcal{G}^{N}_{0}\backslash\{(0)^{\times N}\},\\
\widetilde{\mathcal{G}^{N}_{1}}=\mathcal{G}^{N}_{1},\\
\widetilde{\mathcal{G}^{N}_{2}}=\mathcal{G}^{N}_{2}\backslash\{(2)^{\times N}\},\\
\widetilde{\mathcal{G}^{N}_{3}}=\{(0)^{\times N},(2)^{\times N}\}.\\
\end{cases}\\
&[N]=[2]
\begin{cases}
\widetilde{\mathcal{G}^{N}_{0}}=\mathcal{G}^{N}_{0}\backslash\{(0)^{\times N},(2)^{\times N}\},\\
\widetilde{\mathcal{G}^{N}_{1}}=\mathcal{G}^{N}_{1},\\
\widetilde{\mathcal{G}^{N}_{2}}=\mathcal{G}^{N}_{2},\\
\widetilde{\mathcal{G}^{N}_{3}}=\{(0)^{\times N},(2)^{\times N}\}.\\
\end{cases}\\
&[N]=[3]
\begin{cases}
\widetilde{\mathcal{G}^{N}_{0}}=\mathcal{G}^{N}_{0}\backslash\{(0)^{\times N}\},\\
\widetilde{\mathcal{G}^{N}_{1}}=\mathcal{G}^{N}_{1},\\
\widetilde{\mathcal{G}^{N}_{2}}=\mathcal{G}^{N}_{2}\backslash\{(2)^{\times N}\},\\
\widetilde{\mathcal{G}^{N}_{3}}=\{(0)^{\times N},(2)^{\times N}\}.\\
\end{cases}
\end{aligned}
\end{equation*}

It follows from the Theorem \ref{th-element} Theorem \ref{th-snoges} that the construction is not unique. In this regard, we give another example for $d=4$ according to Table \ref{tab:distri2}.

\textit{Example}~$2$. In $(\mathbb{C}^{4})^{\otimes N}$, the following sets can also be the strongest nonlocal OGESs.
\begin{equation*}
\begin{aligned}
&[N]=[0]
\begin{cases}
\widetilde{\mathcal{G}^{N}_{0}}=\mathcal{G}^{N}_{0}\backslash\{(0)^{\times N},~(2)^{\times N}\},\\
\widetilde{\mathcal{G}^{N}_{1}}=\mathcal{G}^{N}_{1},\\
\widetilde{\mathcal{G}^{N}_{2}}=\mathcal{G}^{N}_{2},\\
\widetilde{\mathcal{G}^{N}_{3}}=\{(0)^{\times N},(2)^{\times N}\}.\\
\end{cases}
\end{aligned}
\end{equation*}
\begin{equation*}
\begin{aligned}
&[N]=[1]
\begin{cases}
\widetilde{\mathcal{G}^{N}_{0}}=\mathcal{G}^{N}_{0}\backslash\{(0)^{\times N}\},\\
\widetilde{\mathcal{G}^{N}_{1}}=\mathcal{G}^{N}_{1}\backslash\{(1)^{\times N}\},\\
\widetilde{\mathcal{G}^{N}_{2}}=\mathcal{G}^{N}_{2},\\
\widetilde{\mathcal{G}^{N}_{3}}=\{(0)^{\times N},(1)^{\times N}\}.\\
\end{cases}\\
&[N]=[2]
\begin{cases}
\widetilde{\mathcal{G}^{N}_{0}}=\mathcal{G}^{N}_{0}\backslash\{(0)^{\times N}\},\\
\widetilde{\mathcal{G}^{N}_{1}}=\mathcal{G}^{N}_{1},\\
\widetilde{\mathcal{G}^{N}_{2}}=\mathcal{G}^{N}_{2}\backslash\{(3)^{\times N}\},\\
\widetilde{\mathcal{G}^{N}_{3}}=\{(0)^{\times N},(3)^{\times N}\}.\\
\end{cases}\\
&[N]=[3]
\begin{cases}
\widetilde{\mathcal{G}^{N}_{0}}=\mathcal{G}^{N}_{0}\backslash\{(0)^{\times N}\},\\
\widetilde{\mathcal{G}^{N}_{1}}=\mathcal{G}^{N}_{1}\backslash\{(3)^{\times N}\},\\
\widetilde{\mathcal{G}^{N}_{2}}=\mathcal{G}^{N}_{2},\\
\widetilde{\mathcal{G}^{N}_{3}}=\{(0)^{\times N},(3)^{\times N}\}.\\
\end{cases}
\end{aligned}
\end{equation*}

The proof follows a similar approach as Theorem \ref{th-snoges} and will not be repeated.
Compared with Ref. \cite{Quantum.7.1101}, the set in our construction contains fewer states as $N$ gradually increases. As shown in Tables \ref{tab:d=4cp}, \ref{tab:d=5cp}, \ref{tab:d=6cp}, \ref{tab:d=7cp}, for $d=4,5$ the size of states in our construction is significantly reduced when $N>4$. For $d=6,7$ our size is much smaller when $N>3$.

\begin{table}[h]
\vspace{-1.0em}
\caption{Comparisons of the sizes of strongest nonlocal OGESs in $(\mathbb{C}^{4})^{\otimes N}$.}
\renewcommand\arraystretch{1.0}
\begin{tabular}{l c  c  c  c  c  c }
 \hline
 \hline
 References & $N=3$ & $N=4$ & $N=5$  & $N=6$ & $N=7$ & $N=8$ \\
 \hline
 Ref. \cite{Quantum.7.1101} & $38$  & $176$ & $782$& $3368$ & $14198$ & $58976$ \\
This work         & $48$  & $192$ & $768$ & 3072 & $12288$ & $49152$ \\
\hline
\hline
\end{tabular}\label{tab:d=4cp}
\end{table}

\begin{table}[h]
\vspace{-2.0em}
\caption{Comparisons of the sizes of strongest nonlocal OGESs in $(\mathbb{C}^{5})^{\otimes N}$.}
\renewcommand\arraystretch{1.0}
\begin{tabular}{l c  c  c  c  c  c }
 \hline
 \hline
 References       & $N=3$ & $N=4$ & $N=5$  & $N=6$ & $N=7$ & $N=8$ \\
 \hline
 Ref. \cite{Quantum.7.1101} & $62$  & $370$ & $2102$& $11530$ & $61742$ & $325090$ \\
 This work         & $75$  & $375$ & $1875$ & $9375$ & $46875$ & $234375$ \\
\hline
\hline
\end{tabular}\label{tab:d=5cp}
\end{table}
\begin{table}[h]
\vspace{-2.0em}
\caption{Comparisons of the sizes of strongest nonlocal OGESs in $(\mathbb{C}^{6})^{\otimes N}$.}
\renewcommand\arraystretch{1.0}
\begin{tabular}{l c  c  c  c  c  c }
 \hline
 \hline
 References                 & $N=3$ & $N=4$ & $N=5$  & $N=6$ & $N=7$ & $N=8$ \\
 \hline
 Ref. \cite{Quantum.7.1101} & $92$  & $672$ & $4652$& $31032$ & $201812$ & $1288992$ \\
 This work                   & $108$  & $648$ & $3888$ & $23328$ & $139968$ & $839808$ \\
\hline
\hline
\end{tabular}\label{tab:d=6cp}
\end{table}
\begin{table}[h]
\vspace{-1.0em}
\caption{Comparisons of the sizes of strongest nonlocal OGESs in $(\mathbb{C}^{7})^{\otimes N}$.}
\renewcommand\arraystretch{1.0}
\begin{tabular}{l c  c  c  c  c  c }
 \hline
 \hline
 References                 & $N=3$ & $N=4$ & $N=5$  & $N=6$ & $N=7$ & $N=8$ \\
 \hline
 Ref. \cite{Quantum.7.1101} & $128$  & $1106$ & $9032$& $70994$ & $543608$ & $4085186$ \\
 This work                  & $147$  & $1029$ & $7203$ & $50421$ & $352947$  & $2470629$ \\
\hline
\hline
\end{tabular}\label{tab:d=7cp}
\end{table}
\vspace{18cm}
\section{Conclusion}\label{Q4}
In this paper, we investigate the strongest nonlocality in $N$-partite quantum system. First, we provide a sufficient and necessary condition for strongest nonlocal sets under some condition. Based on this condition, the minimum size of strongest nonlocal set in system $(\mathbb{C}^{3})^{\otimes N}$ is proven and supported by the OGESs constructed in Ref. \cite{PRA.109.022220}.
Furthermore, we present a general construction in  $(\mathbb{C}^{d})^{\otimes N}$ that demonstrates the existence of smaller strongest nonlocal set with genuine entanglement in high-dimensional multipartite systems. This construction also give an answer to
the question proposed in Ref. \cite{PRA.105.022209}, ``How do we construct a strongly
nonlocal orthogonal genuinely entangled set in $(\mathbb{C}^{d})^{\otimes N}$
for any $d\geq2$ and $N\geq5$?" Our work could enrich the understanding of the strongest nonlocality in multipartite systems.
\vspace{-1.0em}
\section{Acknowledgments}
This work was supported by the National Natural Science Foundation of China under Grants No. 12071110 and No. 62271189, the Hebei Central Guidance on Local Science and Technology Development Foundation of China under Grant No. 236Z7604G.
\appendix
\begin{widetext}
\section{ PROOF OF PROPOSITION~1}\label{A}
Before proving the proposition, we  introduce the following conception and notation.

$Circulant~Matrix$ \cite{D. S.MM.2005}.  A $d\times d$ circulant matrix $B$ is generated from the $d$-dimensional vector $[b_{0},\cdots,b_{d-1}]$ by cyclically permuting its entries, and is
\begin{equation*}
B
=
\left [
\renewcommand\arraystretch{1.1}
\setlength{\arraycolsep}{0.4pt}
\begin{array}{cccccc}
b_{0}~&b_{1}~&b_{2}~&\cdots~&b_{d-2}~&b_{d-1}\\
b_{d-1}~&b_{0}&b_{1}~~&\cdots~&b_{d-3}~&b_{d-2}\\
b_{d-2}~&b_{d-1}&b_{0}~~&\cdots~&b_{d-4}~&b_{d-3}\\
\vdots~&\vdots~&\vdots&\ddots&~\vdots&~\vdots\\
b_{2}~&b_{3}~&b_{4}~&\cdots~&b_{0}~&b_{1}\\
b_{1}~&b_{2}~&b_{3}~&\cdots~&b_{d-1}~&b_{0}
\end{array}
\right].
\end{equation*}
In other word, a circulant matrix is cyclically shifted to the right by one position per row to form the subsequent rows.

Next, we use the circulant matrix to rewrite  Eq. (\ref{eq:GN}). Obviously,
\begin{equation}\label{eq:GNi}
\begin{aligned}
\mathcal{G}^{N}_{i}=&\left(\{i\}\times\mathcal{G}^{N-1}_{0}\right)\bigcup\left(\{i\oplus_{d}(d-1)\}\times\mathcal{G}^{N-1}_{1}\right)\bigcup
\cdots\bigcup\left(\{i\oplus_{d}2\}\times\mathcal{G}^{N-1}_{d-2}\right)\bigcup\left(\{i\oplus_{d}1\}\times\mathcal{G}^{N-1}_{d-1}\right).
\end{aligned}
\end{equation}
We denote (\ref{eq:GNi}) as
\begin{equation}\label{eq:Gni}
\begin{aligned}
&\left(\{i\}\times\mathcal{G}^{N-1}_{0}\right)\bigcup\left(\{i\oplus_{d}(d-1)\}\times\mathcal{G}^{N-1}_{1}\right)\bigcup
\cdots\bigcup\left(\{i\oplus_{d}2\}\times\mathcal{G}^{N-1}_{d-2}\right)\bigcup\left(\{i\oplus_{d}1\}\times\mathcal{G}^{N-1}_{d-1}\right)\\
=&[i,i\oplus_{d}(d-1),\cdots,i\oplus_{d}2,i\oplus_{d}1]
\times\left [
\renewcommand\arraystretch{1.2}
\setlength{\arraycolsep}{0.1pt}
\begin{array}{cccccc}
\mathcal{G}^{N-1}_{0}\\
\mathcal{G}^{N-1}_{1}\\
\vdots\\
\mathcal{G}^{N-1}_{d-2}\\
\mathcal{G}^{N-1}_{d-1}\\
\end{array}
\right ].
\end{aligned}
\end{equation}
For each $i$ in Eq. (\ref{eq:GNi}), one gets
\begin{equation}\label{eq:ZNi}
\begin{aligned}
&\mathcal{G}^{N}_{0}=\left(\{0\}\times\mathcal{G}^{N-1}_{0}\right)\bigcup\left(\{d-1\}\times\mathcal{G}^{N-1}_{1}\right)
\bigcup\cdots\bigcup
\left(\{2\}\times\mathcal{G}^{N-1}_{d-2}\right)\bigcup\left(\{1\}\times\mathcal{G}^{N-1}_{d-1}\right),\\
&\mathcal{G}^{N}_{1}=\left(\{1\}\times\mathcal{G}^{N-1}_{0}\right)\bigcup\left(\{0\}\times\mathcal{G}^{N-1}_{1}\right)
\bigcup\cdots\bigcup
\left(\{3\}\times\mathcal{G}^{N-1}_{d-2}\right)\bigcup\left(\{2\}\times\mathcal{G}^{N-1}_{d-1}\right),\\
&~~~~~~~\vdots\\
&\mathcal{G}^{N}_{d-2}=\left(\{d-2\}\times\mathcal{G}^{N-1}_{0}\right)\bigcup\left(\{d-3\}\times\mathcal{G}^{N-1}_{1}\right)
\bigcup\cdots\bigcup
\left(\{0\}\times\mathcal{G}^{N-1}_{d-2}\right)\bigcup\left(\{d-1\}\times\mathcal{G}^{N-1}_{d-1}\right),\\
&\mathcal{G}^{N}_{d-1}=\left(\{d-1\}\times\mathcal{G}^{N-1}_{0}\right)\bigcup\left(\{d-2\}\times\mathcal{G}^{N-1}_{1}\right)
\bigcup\cdots\bigcup
\left(\{1\}\times\mathcal{G}^{N-1}_{d-2}\right)\bigcup\left(\{0\}\times\mathcal{G}^{N-1}_{d-1}\right),
\end{aligned}
\end{equation}
which can be denoted as
\begin{equation}\label{eq:GNandGN-1}
\left [
\renewcommand\arraystretch{1.1}
\begin{array}{ccc}{}
\mathcal{G}^{N}_{0}\\
\mathcal{G}^{N}_{1}\\
\vdots\\
\mathcal{G}^{N}_{d-2}\\
\mathcal{G}^{N}_{d-1}\\
\end{array}
\right ]=\left [
\renewcommand\arraystretch{1.1}
\setlength{\arraycolsep}{0.4pt}
\begin{array}{ccccc}
0~&d-1~&\cdots~&2~&1\\
1~&0~&\cdots~&3~&2\\
\vdots~&\vdots~&\ddots&~\vdots&~\vdots\\
d-2~&d-3~&\cdots~&0~&d-1\\
d-1~&d-2~&\cdots~&1~&0
\end{array}
\right]
\times
\left [
\renewcommand\arraystretch{1.1}
\begin{array}{ccc}{}
\mathcal{G}^{N-1}_{0}\\
\mathcal{G}^{N-1}_{1}\\
\vdots\\
\mathcal{G}^{N-1}_{d-2}\\
\mathcal{G}^{N-1}_{d-1}\\
\end{array}
\right ].
\end{equation}
Here $
\left [
\renewcommand\arraystretch{1.1}
\setlength{\arraycolsep}{0.4pt}
\begin{array}{ccccc}
0~&d-1~&\cdots~&2~&1\\
1~&0~&\cdots~&3~&2\\
\vdots~&\vdots~&\ddots&~\vdots&~\vdots\\
d-2~&d-3~&\cdots~&0~&d-1\\
d-1~&d-2~&\cdots~&1~&0
\end{array}
\right]$
is a circulant matrix that generated from the $d$-dimensional vector $[0,d-1,\cdots,2,1]$.

Later on, we will give the proof of Proposition \ref{pr1}. Obviously, the sets $\mathcal{G}^{1}_{0},\cdots,\mathcal{G}^{1}_{d-1}$ are pairwise disjoint and $\bigcup^{d-1}_{i=0}\mathcal{G}^{1}_{i}=\mathbb{Z}_{d}$. We prove this proposition by induction.
Assume that the claim is true for $N=k$, i.e., $\bigcup^{d-1}_{i=0}\mathcal{G}^{k}_{i}=\mathbb{Z}^{k}_{d}$ and $\mathcal{G}^{k}_{i}\cap\mathcal{G}^{k}_{j}=\emptyset$, if $i\neq j\in\mathbb{Z}_{d}$. Let $l=k+1$, for any two different sets $\mathcal{G}^{l}_{i}=\bigcup^{d-1}_{x=0}\left(\{i_{x}\}\times\mathcal{G}^{k}_{x}\right)$ and $\mathcal{G}^{l}_{j}=\bigcup^{d-1}_{y=0}\left(\{j_{y}\}\times\mathcal{G}^{k}_{y}\right)$, according to Eq. (\ref{eq:GNandGN-1}),
if $i_{x}=i_{y}$, one gets $x\neq y$, from the induction hypothesis, $\mathcal{G}^{k}_{x}\cap\mathcal{G}^{k}_{y}=\emptyset$ holds for $k$, thus $\left(\{i_{x}\}\times\mathcal{G}^{k}_{x}\right)\bigcap\left(\{j_{y}\}\times\mathcal{G}^{k}_{y}\right)=\emptyset$.
If $i_{x}\neq j_{y}$, it definitely lead to $\left(\{i_{x}\}\times\mathcal{G}^{k}_{x}\right)\bigcap\left(\{j_{y}\}\times\mathcal{G}^{k}_{y}\right)=\emptyset$.
In conclusion, $\mathcal{G}^{l}_{i}\cap\mathcal{G}^{l}_{j}=\emptyset$.

On the other hand,
\begin{equation*}
\begin{aligned}
\bigcup^{d-1}_{i=0}\mathcal{G}^{l}_{i}&=\bigcup^{d-1}_{i=0}\bigcup^{d-1}_{j=0}\left(\{i\oplus_{d}(d-j)\}\times\mathcal{G}^{k}_{j}\right)\\
&=\bigcup^{d-1}_{j=0}\left(\left(\bigcup^{d-1}_{i=0}\{i\oplus_{d}(d-j)\}\right)\times\mathcal{G}^{k}_{j}\right)\\
&=\bigcup^{d-1}_{j=0}\left(\{0,1,\cdots,d-1\}\times\mathcal{G}^{k}_{j}\right)\\
&=\{0,1,\cdots,d-1\}\times\bigcup^{d-1}_{j=0}\mathcal{G}^{k}_{j}\\
&=\{0,1,\cdots,d-1\}\times \mathbb{Z}^{k}_{d}\\
&=\mathbb{Z}^{l}_{d}.
\end{aligned}
\end{equation*}
The proof is completed. $\hfill\blacksquare$
\section{PROOF OF PROPOSITION~2}\label{B}

Before the proof, we first give the following notation.

Assume that $P_{d}=\{\sigma^{r}|r\in \mathbb{Z}_{d}\}$ is a cyclic permutation group of order $d$, and
\begin{equation*}
\sigma[i_{0},i_{1},\cdots,i_{d-1}]=[i_{d-1},i_{0},\cdots,i_{d-2}],
\end{equation*}
where $[i_{0},i_{1},\cdots,i_{d-1}]$ is an arbitrary $d$-dimensional vector.

The proof of Proposition \ref{pr2} is as follows. Observing Eq.\;(\ref{eq:Gni}), we have that shifting the component of $[i,i\oplus_{d}(d-1),\cdots,i\oplus_{d}2,i\oplus_{d}1]$ to the left by one position is equivalent to shifting the component of $\left [
\renewcommand\arraystretch{1.2}
\setlength{\arraycolsep}{0.1pt}
\begin{array}{cccccc}
\mathcal{G}^{N-1}_{0}\\
\mathcal{G}^{N-1}_{1}\\
\vdots\\
\mathcal{G}^{N-1}_{d-2}\\
\mathcal{G}^{N-1}_{d-1}\\
\end{array}
\right ]$ to down by one position, that is,
\begin{equation}\label{eq:sigmal-1}
\begin{aligned}
&\sigma^{-1}[i,i\oplus_{d}(d-1),\cdots,i\oplus_{d}2,i\oplus_{d}1]
\times\left [
\renewcommand\arraystretch{1.2}
\setlength{\arraycolsep}{0.1pt}
\begin{array}{cccccc}
\mathcal{G}^{N-1}_{0}\\
\mathcal{G}^{N-1}_{1}\\
\vdots\\
\mathcal{G}^{N-1}_{d-2}\\
\mathcal{G}^{N-1}_{d-1}\\
\end{array}
\right ]
=
[i,i\oplus_{d}(d-1),\cdots,i\oplus_{d}2,i\oplus_{d}1]
\times\sigma\left [
\renewcommand\arraystretch{1.2}
\setlength{\arraycolsep}{0.1pt}
\begin{array}{cccccc}
\mathcal{G}^{N-1}_{0}\\
\mathcal{G}^{N-1}_{1}\\
\vdots\\
\mathcal{G}^{N-1}_{d-2}\\
\mathcal{G}^{N-1}_{d-1}\\
\end{array}
\right ].
\end{aligned}
\end{equation}
In Eq. (\ref{eq:GNandGN-1}),
performing a permutation $\sigma^{-1}$ for each row of $\left [
\renewcommand\arraystretch{1.1}
\setlength{\arraycolsep}{0.4pt}
\begin{array}{ccccc}
0~&d-1~&\cdots~&2~&1\\
1~&0~&\cdots~&3~&2\\
\vdots~&\vdots~&\ddots&~\vdots&~\vdots\\
d-2~&d-3~&\cdots~&0~&d-1\\
d-1~&d-2~&\cdots~&1~&0
\end{array}
\right]$, one gets
\begin{equation}\label{eq:M=M}
\begin{aligned}
\left [
\renewcommand\arraystretch{1.1}
\setlength{\arraycolsep}{0.4pt}
\begin{array}{ccccc}
d-1~&d-2~&\cdots~&1~&0\\
0~&d-1~&\cdots~&2~&1\\
\vdots~&\vdots~&\ddots&~\vdots&~\vdots\\
d-3~&d-4~&\cdots~&d-1~&d-2\\
d-2~&d-3~&\cdots~&0~&d-1
\end{array}
\right]
\times
\left [
\renewcommand\arraystretch{1.1}
\begin{array}{ccc}{}
\mathcal{G}^{N-1}_{0}\\
\mathcal{G}^{N-1}_{1}\\
\vdots\\
\mathcal{G}^{N-1}_{d-2}\\
\mathcal{G}^{N-1}_{d-1}\\
\end{array}
\right ]
=\left [
\renewcommand\arraystretch{1.1}
\setlength{\arraycolsep}{0.4pt}
\begin{array}{ccccc}
0~&d-1~&\cdots~&2~&1\\
1~&0~&\cdots~&3~&2\\
\vdots~&\vdots~&\ddots&~\vdots&~\vdots\\
d-2~&d-3~&\cdots~&0~&d-1\\
d-1~&d-2~&\cdots~&1~&0
\end{array}
\right]
\times
\sigma\left [
\renewcommand\arraystretch{1.1}
\begin{array}{ccc}{}
\mathcal{G}^{N-1}_{0}\\
\mathcal{G}^{N-1}_{1}\\
\vdots\\
\mathcal{G}^{N-1}_{d-2}\\
\mathcal{G}^{N-1}_{d-1}\\
\end{array}
\right ].
\end{aligned}
\end{equation}
On the other hand, one has
\begin{equation}\label{eq:sigmaVN}
\begin{aligned}
&\left [
\renewcommand\arraystretch{1.1}
\setlength{\arraycolsep}{0.4pt}
\begin{array}{ccccc}
d-1~&d-2~&\cdots~&1~&0\\
0~&d-1~&\cdots~&2~&1\\
\vdots~&\vdots~&\ddots&~\vdots&~\vdots\\
d-3~&d-4~&\cdots~&d-1~&d-2\\
d-2~&d-3~&\cdots~&0~&d-1
\end{array}
\right]
\times
\left [
\renewcommand\arraystretch{1.1}
\begin{array}{ccc}{}
\mathcal{G}^{N-1}_{0}\\
\mathcal{G}^{N-1}_{1}\\
\vdots\\
\mathcal{G}^{N-1}_{d-2}\\
\mathcal{G}^{N-1}_{d-1}\\
\end{array}
\right ]
=\left [
\renewcommand\arraystretch{1.1}
\begin{array}{ccc}{}
\mathcal{G}^{N}_{d-1}\\
\mathcal{G}^{N}_{0}\\
\vdots\\
\mathcal{G}^{N}_{d-3}\\
\mathcal{G}^{N}_{d-2}\\
\end{array}
\right ]
=\sigma\left [
\renewcommand\arraystretch{1.1}
\begin{array}{ccc}{}
\mathcal{G}^{N}_{0}\\
\mathcal{G}^{N}_{1}\\
\vdots\\
\mathcal{G}^{N}_{d-2}\\
\mathcal{G}^{N}_{d-1}\\
\end{array}
\right ].
\end{aligned}
\end{equation}
By Eq. (\ref{eq:M=M}) and Eq. (\ref{eq:sigmaVN}) we obtain
\begin{equation}
\begin{aligned}
\sigma\left [
\renewcommand\arraystretch{1.1}
\begin{array}{ccc}{}
\mathcal{G}^{N}_{0}\\
\mathcal{G}^{N}_{1}\\
\vdots\\
\mathcal{G}^{N}_{d-2}\\
\mathcal{G}^{N}_{d-1}\\
\end{array}
\right ]=
\left [
\renewcommand\arraystretch{1.1}
\setlength{\arraycolsep}{0.4pt}
\begin{array}{ccccc}
0~&d-1~&\cdots~&2~&1\\
1~&0~&\cdots~&3~&2\\
\vdots~&\vdots~&\ddots&~\vdots&~\vdots\\
d-2~&d-3~&\cdots~&0~&d-1\\
d-1~&d-2~&\cdots~&1~&0
\end{array}
\right]
\times
\sigma\left [
\renewcommand\arraystretch{1.1}
\begin{array}{ccc}{}
\mathcal{G}^{N-1}_{0}\\
\mathcal{G}^{N-1}_{1}\\
\vdots\\
\mathcal{G}^{N-1}_{d-2}\\
\mathcal{G}^{N-1}_{d-1}\\
\end{array}
\right ].
\end{aligned}
\end{equation}

Let
\begin{equation}\label{eq:VN}
\mathcal{V}_{N}:= \left [
\renewcommand\arraystretch{1.1}
\begin{array}{ccc}{}
\mathcal{G}^{N}_{0}\\
\mathcal{G}^{N}_{1}\\
\vdots\\
\mathcal{G}^{N}_{d-2}\\
\mathcal{G}^{N}_{d-1}\\
\end{array}
\right ]
\end{equation}
and
\begin{equation}\label{eq:Md}
\mathcal{M}_{d}:=\left [
\renewcommand\arraystretch{1.1}
\setlength{\arraycolsep}{0.4pt}
\begin{array}{ccccc}
0~&d-1~&\cdots~&2~&1\\
1~&0~&\cdots~&3~&2\\
\vdots~&\vdots~&\ddots&~\vdots&~\vdots\\
d-2~&d-3~&\cdots~&0~&d-1\\
d-1~&d-2~&\cdots~&1~&0
\end{array}
\right],
\end{equation}
we get
\begin{equation}\label{eq:VN,VN-1}
\sigma
\mathcal{V}_{N}=\mathcal{M}_{d}\times
\sigma\mathcal{V}_{N-1}.
\end{equation}
Generating a circulant matrix of order $d$ with column vectors $\mathcal{V}_{N}$, $\sigma\mathcal{V}_{N}$,$\cdots$,$\sigma^{(d-1)}\mathcal{V}_{N}$, by Eq. (\ref{eq:VN,VN-1}), there is
\begin{equation}\label{eq:M3}
\setlength{\arraycolsep}{0.2pt}
\small{
\begin{aligned}
&\left [
\begin{array}{ccccc}
\mathcal{V}_{N},&\sigma\mathcal{V}_{N},\cdots,&\sigma^{(d-2)}\mathcal{V}_{N},&\sigma^{(d-1)}\mathcal{V}_{N}\\
\end{array}
\right ]\\
=
&\mathcal{M}_{d}\times
\left [
\renewcommand\arraystretch{1.1}
\setlength{\arraycolsep}{0.2pt}
\begin{array}{cccccc}
\mathcal{V}_{N-1},&\sigma\mathcal{V}_{N-1},\cdots,&\sigma^{(d-2)}\mathcal{V}_{N-1},&\sigma^{(d-1)}\mathcal{V}_{N-1}
\end{array}
\right ].
\end{aligned}
}
\end{equation}
Repeating this argument, we obatin
\begin{equation}\label{eq:NMd}
\setlength{\arraycolsep}{0.2pt}
\small{
\begin{aligned}
&\left [
\begin{array}{ccccc}
\mathcal{V}_{N},&\sigma\mathcal{V}_{N},\cdots,&\sigma^{(d-2)}\mathcal{V}_{N},&\sigma^{(d-1)}\mathcal{V}_{N}\\
\end{array}
\right ]\\
=
&\underbrace{\mathcal{M}_{d}\times\cdots\times \mathcal{M}_{d}}_{N-1}
\times
\left [
\renewcommand\arraystretch{1.1}
\setlength{\arraycolsep}{0.2pt}
\begin{array}{cccccc}
\mathcal{V}_{1},&\sigma\mathcal{V}_{1},\cdots,&\sigma^{(d-2)}\mathcal{V}_{1},&\sigma^{(d-1)}\mathcal{V}_{1}
\end{array}
\right ]\\
=
&\underbrace{\mathcal{M}_{d}\times\cdots\times \mathcal{M}_{d}}_{N-1}\times \mathcal{M}_{d}.
\end{aligned}
}
\end{equation}
Here we use  $[\mathcal{V}_{1},\sigma\mathcal{V}_{1},\cdots,\sigma^{(d-2)}\mathcal{V}_{1},\sigma^{(d-1)}\mathcal{V}_{1}]=\mathcal{M}_{d}$. Substituting Eq. (\ref{eq:VN}) and Eq. (\ref{eq:Md}) into Eq. (\ref{eq:NMd}), we have
\begin{equation}\label{eq:GNoder}
\renewcommand\arraystretch{1.1}
\setlength{\arraycolsep}{1pt}
\small{
\begin{aligned}
\left [
\begin{array}{ccccc}
\mathcal{G}^{N}_{0}&\mathcal{G}^{N}_{d-1}&~\cdots~&\mathcal{G}^{N}_{2}&\mathcal{G}^{N}_{1}\\
\mathcal{G}^{N}_{1}&\mathcal{G}^{N}_{0}&~\cdots~&\mathcal{G}^{N}_{3}&\mathcal{G}^{N}_{2}\\
\vdots&\vdots&\ddots&\vdots&\vdots\\
\mathcal{G}^{N}_{d-2}&\mathcal{G}^{N}_{d-3}&~\cdots~&\mathcal{G}^{N}_{0}&\mathcal{G}^{N}_{d-1}\\
\mathcal{G}^{N}_{d-1}&\mathcal{G}^{N}_{d-2}&~\cdots~&\mathcal{G}^{N}_{1}&\mathcal{G}^{N}_{0}\\
\end{array}
\right]
=\left [
\renewcommand\arraystretch{1.1}
\setlength{\arraycolsep}{0.4pt}
\begin{array}{ccccc}
0~&d-1~&\cdots~&2~&1\\
1~&0~&\cdots~&3~&2\\
\vdots~&\vdots~&\ddots&~\vdots&~\vdots\\
d-2~&d-3~&\cdots~&0~&d-1\\
d-1~&d-2~&\cdots~&1~&0
\end{array}
\right]_{N}\times
\left [
\renewcommand\arraystretch{1.1}
\setlength{\arraycolsep}{0.4pt}
\begin{array}{ccccc}
0~&d-1~&\cdots~&2~&1\\
1~&0~&\cdots~&3~&2\\
\vdots~&\vdots~&\ddots&~\vdots&~\vdots\\
d-2~&d-3~&\cdots~&0~&d-1\\
d-1~&d-2~&\cdots~&1~&0
\end{array}
\right]_{N-1}\times\cdots\times
\left [
\renewcommand\arraystretch{1.1}
\setlength{\arraycolsep}{0.4pt}
\begin{array}{ccccc}
0~&d-1~&\cdots~&2~&1\\
1~&0~&\cdots~&3~&2\\
\vdots~&\vdots~&\ddots&~\vdots&~\vdots\\
d-2~&d-3~&\cdots~&0~&d-1\\
d-1~&d-2~&\cdots~&1~&0
\end{array}
\right]_{1}
.
\end{aligned}
}
\end{equation}

Beginning at the first square matrix on the right, like a snowball, the set of $N$-tuples is generated. As the right square matrices are the same, no matter how we change the position of the matrices, nothing is going to change on the left square matrix.
Therefore the proof is now complete. $\hfill\blacksquare$

\end{widetext}

\end{document}